\documentclass[runningheads]{llncs}
\usepackage{graphicx}

\usepackage{tikz}
\usepackage{comment}
\usepackage{amsmath,amssymb} 
\usepackage{color}
\usepackage{subfigure}
\usepackage[colorlinks]{hyperref}
\usepackage{amsfonts}
\usepackage{amsmath}
\usepackage{bm}
\usepackage{multirow}
\usepackage{algorithm}
\usepackage{algpseudocode}
\usepackage{booktabs}
\usepackage{bbding}
\usepackage{setspace}

\usepackage[accsupp]{axessibility}  

\usepackage[width=122mm,left=12mm,paperwidth=146mm,height=193mm,top=12mm,paperheight=217mm]{geometry}

\begin{document}
\pagestyle{headings}
\mainmatter
\def\ECCVSubNumber{xxx}  

\title{Dynamic Dual Trainable Bounds for Ultra-low Precision Super-Resolution Networks} 

\titlerunning{Dynamic Dual Trainable Bounds (DDTB)}
\authorrunning{Yunshan Zhong \emph{et al}.}
%
\author{Yunshan Zhong$^1$, Mingbao Lin$^{2,3}$, Xunchao Li$^2$, Ke Li$^3$,\\ Yunhang Shen$^3$, Fei Chao$^{1,2}$, Yongjian Wu$^3$, Rongrong Ji$^{1,2}$\thanks{Corresponding Author}}
\institute{$^1$Institute of Artificial Intelligence, Xiamen University\\
$^2$MAC Lab, Department of Artificial Intelligence, Xiamen University\\
$^3$Tencent Youtu Lab \\
{\tt\small \{zhongyunshan, lmbxmu, lixunchao\}@stu.xmu.edu.cn,}\\ 
{\tt\small \{tristanli.sh, shenyunhang01\}@gmail.com,}\\
{\tt\small fchao@xmu.edu.cn,}
{\tt\small littlekenwu@tencent.com,}
{\tt\small rrji@xmu.edu.cn}}
\maketitle
\vspace{-2em}
\begin{abstract}
Light-weight super-resolution (SR) models have received considerable attention for their serviceability in mobile devices. Many efforts employ network quantization to compress SR models. However, these methods suffer from severe performance degradation when quantizing the SR models to ultra-low precision (\emph{e.g.}, 2-bit and 3-bit) with the low-cost layer-wise quantizer. 
In this paper, we identify that the performance drop comes from the contradiction between the layer-wise symmetric quantizer and the highly asymmetric activation distribution in SR models. This discrepancy leads to either a waste on the quantization levels or detail loss in reconstructed images. 
Therefore, we propose a novel activation quantizer, referred to as Dynamic Dual Trainable Bounds (DDTB), to accommodate the asymmetry of the activations. 
Specifically, DDTB innovates in: 
1) A layer-wise quantizer with trainable upper and lower bounds to tackle the highly asymmetric activations. 
2) A dynamic gate controller to adaptively adjust the upper and lower bounds at runtime to overcome the drastically varying activation ranges over different samples.
To reduce the extra overhead, the dynamic gate controller is quantized to 2-bit and applied to only part of the SR networks according to the introduced dynamic intensity.
Extensive experiments demonstrate that our DDTB exhibits significant performance improvements in ultra-low precision. For example, our DDTB achieves a 0.70dB PSNR increase on Urban100 benchmark when quantizing EDSR to 2-bit and scaling up output images to $\times$4. Code is at \url{https://github.com/zysxmu/DDTB}.
\vspace{-1.3em}

\end{abstract}

\section{Introduction}

Single image super-resolution (SISR) is a classic yet challenging research topic in low-level computer vision. It aims to construct a high-resolution (HR) image from a given low-resolution (LR) image. Recent years have witnessed the revolution of deep convolutional neural networks (DCNN), which leads to many state-of-the-arts~\cite{lim2017enhanced,zhang2018residual,dong2014learning} in SISR task.

\begin{figure*}[!t]
\centering
\includegraphics[height=0.5\linewidth]{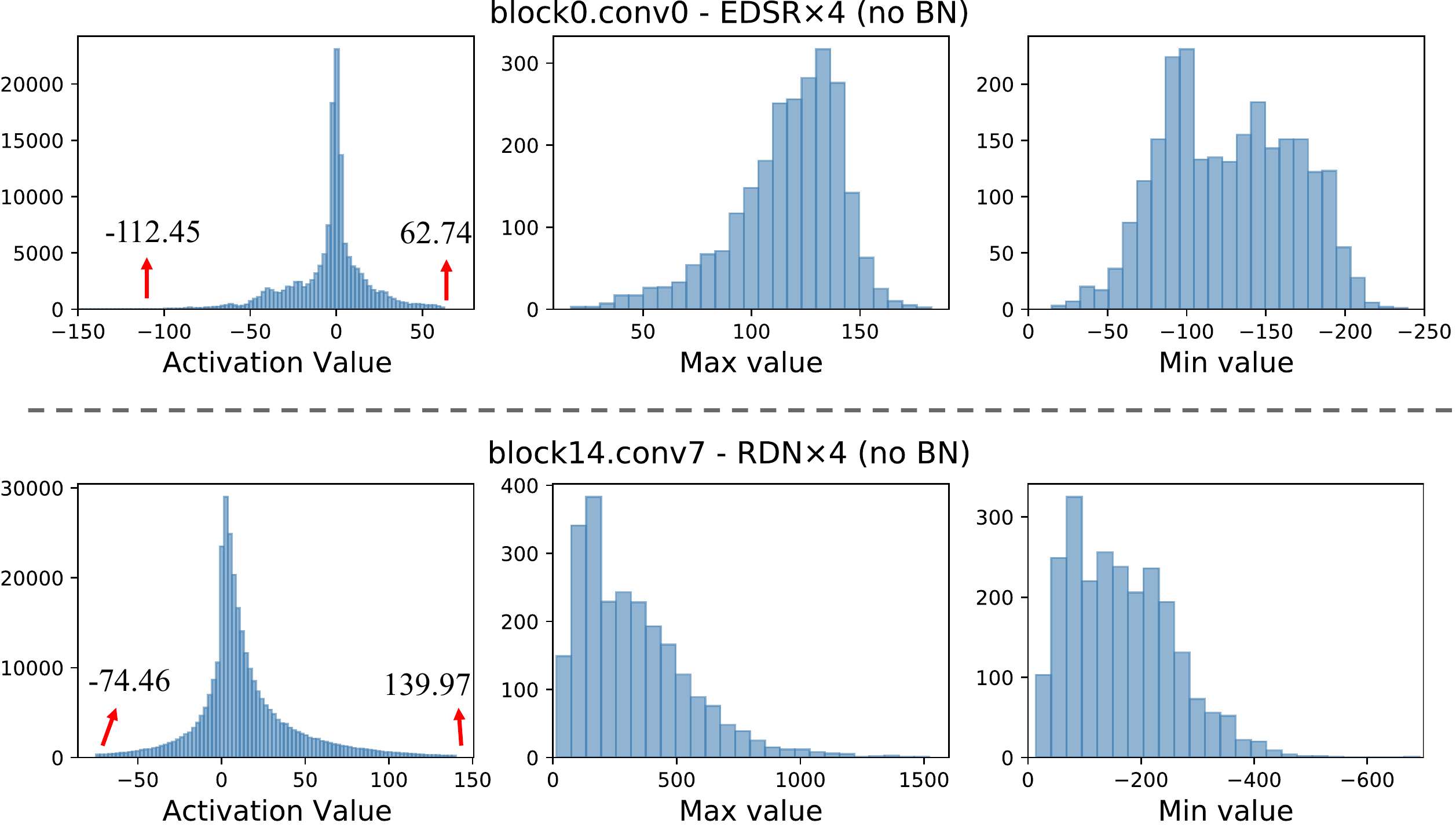}
\vspace{-1.0em}
\caption{The first column shows the activation histograms. The second and third columns show the maximum and minimum activation values of different samples. We perform experiments with EDSR~\cite{lim2017enhanced} and RDN~\cite{zhang2018residual} on DIV2K~\cite{timofte2017ntire} dataset.}
\label{visualization}
\vspace{-1.5em}
\end{figure*}

When looking back on the development of DCNN in SISR, we find that the record-breaking performance is accompanied by a drastically increasing model complexity. SRCNN~\cite{dong2014learning}, the first work to integrate DCNN to SR, has only three convolutional layers with a total of 57K parameters. Then, EDSR~\cite{lim2017enhanced} constructs a 64-layer CNN with 1.5M parameters. Equipped with a residual dense block, RDN~\cite{zhang2018residual} introduces 151 convolutional layers with 22M parameters. Also, it requires around 5,896G float-point operations (FLOPs) to produce only one 1920$\times$1080 image (upscaling factor $\times$4). On the one hand, the high memory footprint and computation cost of DCNN-based SR models barricade their deployment on many resource-hungry platforms such as smartphones, wearable gadgets, embedding devices, \emph{etc}.
On the other hand, SR is particularly popular on these devices where the photograph resolution must be enhanced after being taken by the users.
Therefore, compressing DCNN-based SR models has gained considerable attention from both academia and industries.
In recent years, a variety of methodologies are explored to realize practical deployment~\cite{lin2020hrank,krishnamoorthi2018quantizing,hinton2015distilling,han2015learning}.

By discretizing the full-precision weights and activations within the DCNN, network quantization has emerged as one of the most promising technologies. It reduces not only memory storage for lower-precision representation but computation cost for more efficient integer operations. 
Earlier studies mostly focus on high-level vision tasks, such as classification~\cite{krishnamoorthi2018quantizing,esser2019learned,rastegari2016xnor,lin2020rotated} and segmentation~\cite{xu2018quantization,zhang2021medq}. A direct extension of these methods to SR networks has been proved infeasible since low-vision networks often have different operators with these high-level networks~\cite{li2020pams}. 
Consequently, excavating specialized quantization methods for DCNN-based SR models recently has aroused increasing attention in the research community.
For example, PAMS~\cite{li2020pams} designs a layer-wise quantizer with a learnable clipping to tackle the large ranges of activations, but severe performance degradation occurs in ultra-low precision settings (\emph{e.g.}, 2-bit and 3-bit) as shown in Sec.\,\ref{sec:experimental results}. 
A recent study DAQ~\cite{hong2022daq} adopts a channel-wise distribution-aware quantization scheme. Despite the progress, the performance improvement comes at the cost of considerable overhead from normalizing and de-normalizing feature maps, as well as the expensive per-channel quantizer.
Therefore, existing studies are stuck in either heavy extra costs or severe performance drops when performing ultra-low precision quantization.

\begin{figure*}[!t]
\centering
\subfigure[]{
\includegraphics[width=0.3\linewidth]{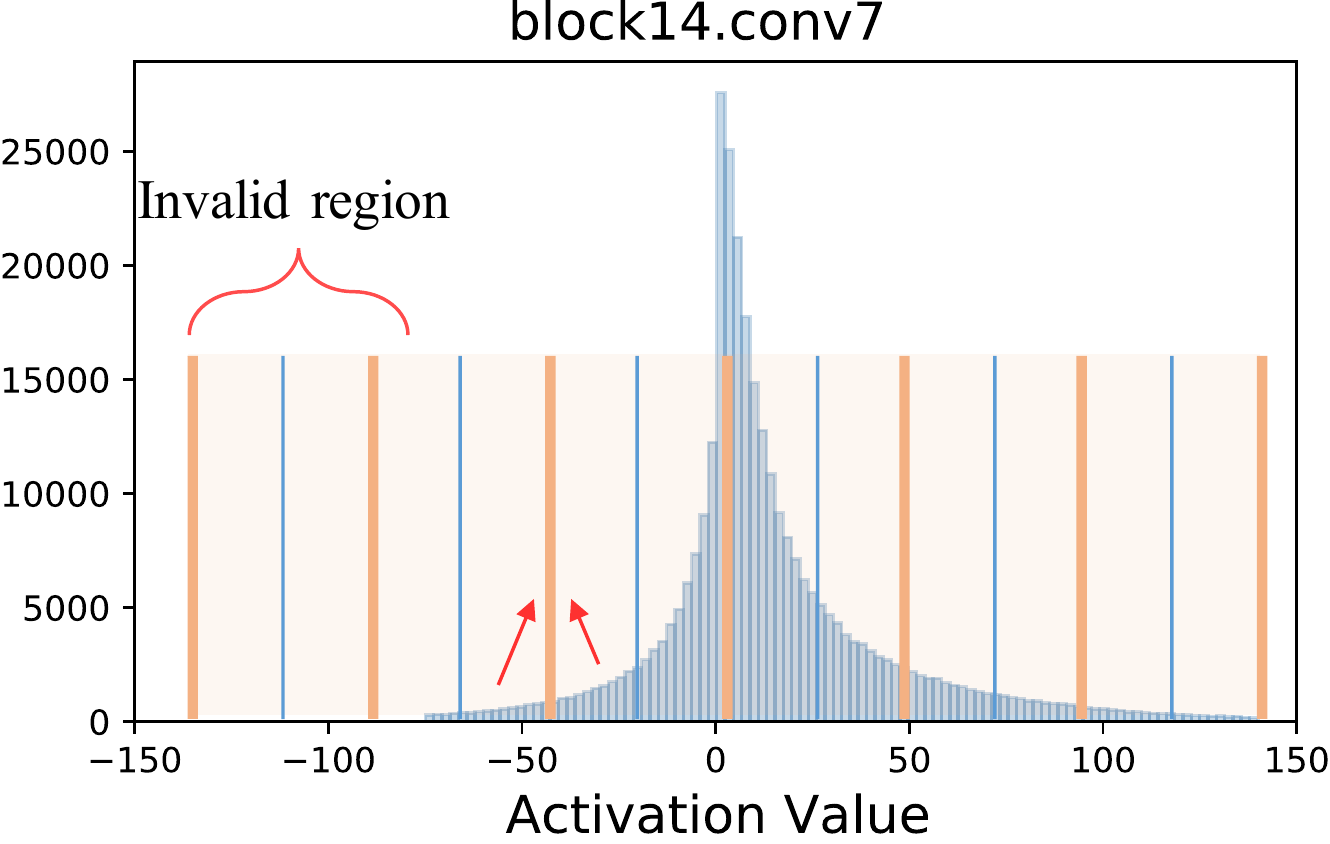} \label{problem:a}
}
\subfigure[]{
\includegraphics[width=0.3\linewidth]{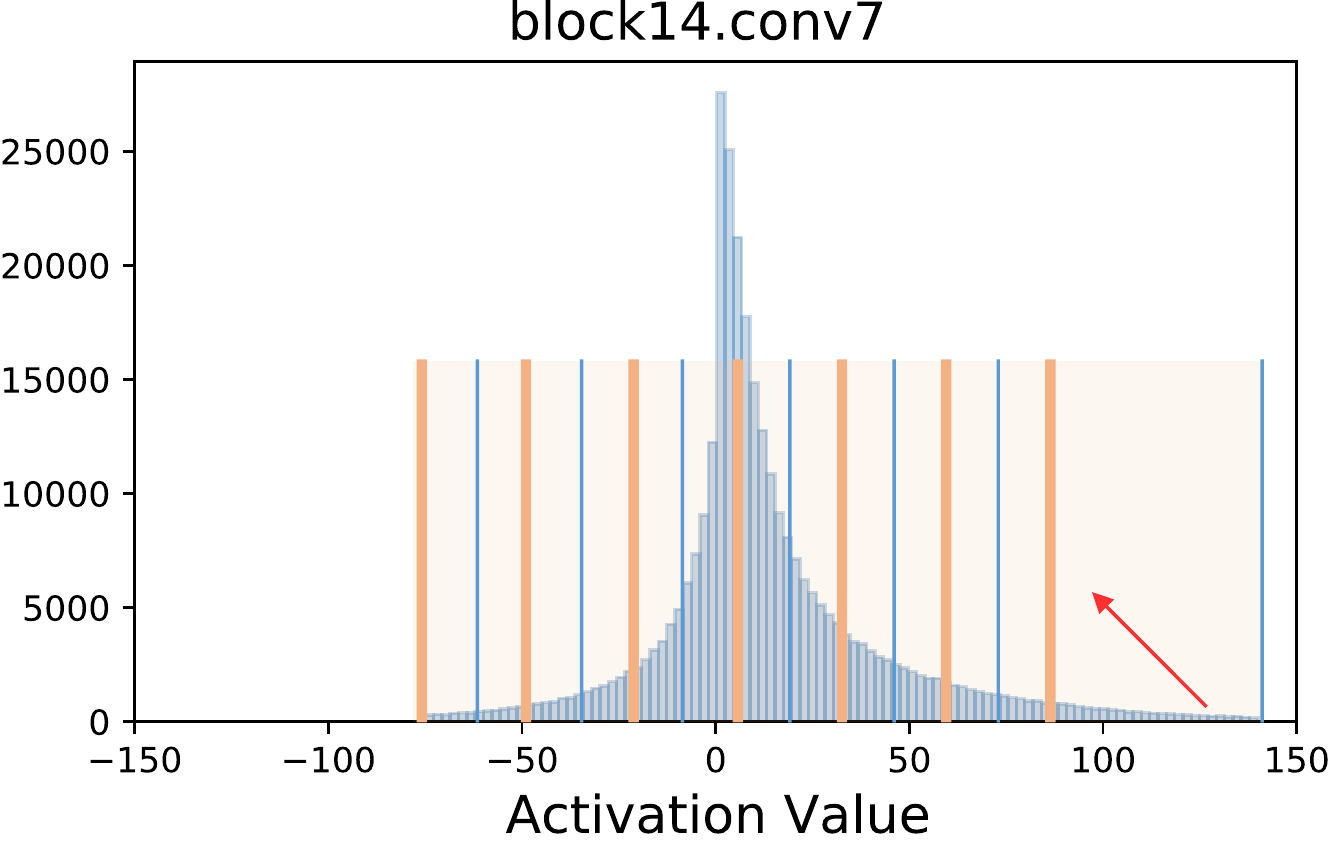}
\label{problem:b} 
}
\subfigure[]{
\includegraphics[width=0.3\linewidth]{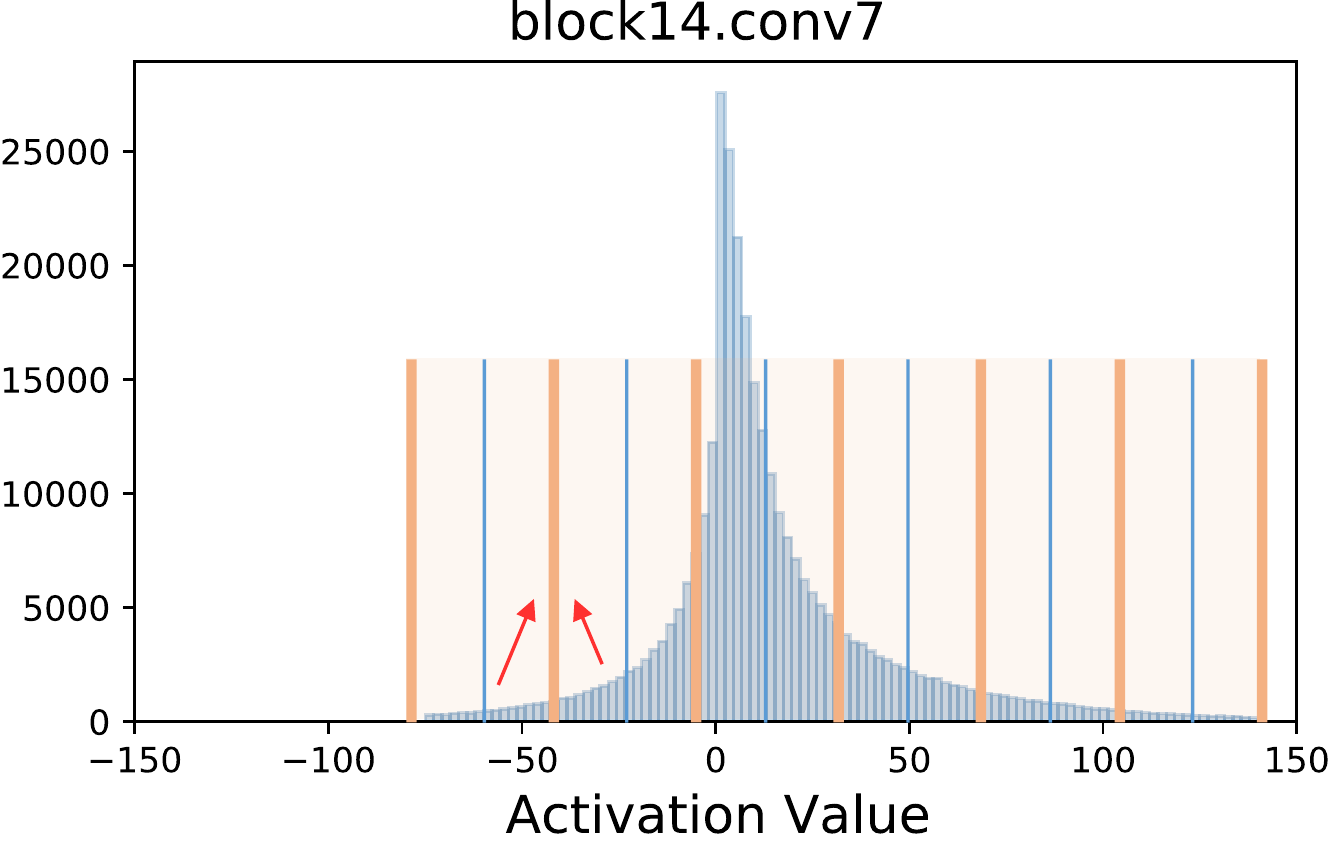}
\label{problem:dual} 
}
\vspace{-1.0em}
\caption{Example of ``quantization unfitness'' using RDN~\cite{zhang2018residual} on DIV2K~\cite{timofte2017ntire}. The orange bar denotes the quantization levels. The red arrows show the quantization level of activations. (a) Two quantization levels are wasted on the region without activations. (b) High-magnitude activations are quantized to a small quantization level. (c) The distribution of the quantization levels from dual bounds in our DDTB.}
\label{problem}
\vspace{-1.5em}
\end{figure*}

In this paper, we realize that the inapplicability of existing studies comes from the contradiction between the layer-wise symmetric quantizer and the asymmetric activation distribution in DCNN-based SR models. 
Specifically, it has been a wide consensus~\cite{lim2017enhanced,zhang2018residual,zhang2018image,haris2018deep} that removing batch normalization (BN) layers increases the super-resolution performance since low-level vision is sensitive to the scale information of images while BN reduces the range flexibility of activations. However, the removal of BN leads to highly asymmetric activation distribution, as well as diverse maximum and minimum activations for different input samples. An illustrative example with EDSR~\cite{lim2017enhanced}, RDN~\cite{zhang2018residual} on DIV2K~\cite{timofte2017ntire} dataset is given in Fig.\,\ref{visualization}. Note that many previous works equip SR networks with BN layers, which, however, have to compensate for the performance drops by other strategies. For example, SRResNet~\cite{ledig2017photo} constructs a highway that propagates the activations of the first convolutional layer to the outputs of every other block, which again leads to activation distributions like Fig.\,\ref{visualization}. 

Although these abnormal activation distributions benefit a full-precision SR network, they are unfriendly to the quantized version which often constructs a symmetric quantizer with only one clipping bound to perform network quantization, resulting in the issue of ``quantization unfitness''. Fig.\,\ref{problem} illustrates a toy example. With an asymmetric activation distribution, a large clipping wastes two quantization levels on the region without any activation items, while a small clipping quantizes large-magnitude activations to a small quantization level which leads to large quantization error and causes details loss in high-resolution images. Thus, an adaptive quantizer to the activation distribution is crucial to the quality of the reconstructed high-resolution images.

Motivated by the above analysis, in this paper, we propose dynamic dual trainable bounds (DDTB) to quantize the activations in SR models. 
Our DDTB innovates in: 1) A layer-wise quantizer with trainable upper and lower bounds to tackle the highly asymmetric activations; 2) A dynamic gate controller to adaptively adjust the upper and lower bounds based on different input samples at runtime for overcoming the drastically varying activation ranges over different samples. 
To minimize the extra costs on computation and storage, the dynamic gate is represented in a 2-bit format and applied to part of the network according to the dynamic intensity in each layer. 
Also, we introduce an initializer to provide a warmup for DDTB. Specifically, we first use the activation statistics from the full-precision network instead of the quantized version to initialize the upper and lower bounds. Then, the dynamic gate controller is trained individually towards its target output of 1 for all inputs in the early several epochs.
Combing with the proposed initializer, DDTB provides significant performance improvements, especially when SR models are quantized to the case of ultra-low bits. For instance, compared with the state-of-the-art, our DDTB achieves performance gains by 0.70dB PSNR on Urban100 benchmark~\cite{huang2015single} when EDSR~\cite{lim2017enhanced} is quantized to 2-bit and outputs $\times$4 reconstructed images.

\section{Related Work}

\subsection{Single Image Super Resolution}
Owing to the strength of deep convolutional neural networks, DCNN-based SR methods have gained great performance boosts and dominated the field of SISR.
SRCNN~\cite{dong2014learning} is the first to construct an end-to-end CNN-based mapping between the low- and high-resolution images. By increasing the network depth, VDSR~\cite{kim2016accurate} manifests significant performance improvements. Nevertheless, the increasing depth also weakens the capacity of overcoming the gradient vanishing problem and retaining image detail. Consequently, the skip-connection based blocks, such as the residual block~\cite{ledig2017photo} and the dense block~\cite{tong2017image}, have become a basic component of the SR models~\cite{lim2017enhanced,zhang2018residual}. Also, many other complex network structures like channel attention mechanism~\cite{zhang2018image,magid2021dynamic} and non-local attention~\cite{mei2020non-local,mei2021non-local} are also integrated into SR for a better performance.

With the increasing popularity of CNNs on resource-hungry devices, developing efficient SR models has aroused much attention recently. Most works in this area are indicated to devising lightweight network architectures.
For example, DRCN~\cite{kim2016deeply} and DRRN~\cite{tai2017image} adopt the recursive structure to increase the network depth while mitigating the model parameters. 
To escape from the expensive up-sampling operator, FSRCNN~\cite{dong2016accelerating} introduces a de-convolutional layer while ESPCN~\cite{shi2016real} devises a sub-pixel convolution module. Many others resort to enhance the efficiency of intermediate feature representation~\cite{lai2017deep,ahn2018fast,hui2019lightweight,luo2020latticenet}.
Nevertheless, these computation savings are very limited compared to costs from the full-precision convolution.

\subsection{Quantized SR Models}


As a promising technique to compress SR models, network quantization has received ever-growing attention~\cite{ma2019efficient,xin2020binarized,jiang2021training,hong2022daq,li2020pams,wang2021fully}. 
Ma~\emph{et al.}~\cite{ma2019efficient} pioneered 1-bit quantization over the weights of SR model. Performance drops severely if binarizing the activations as well. To remedy this issue, the structure of SR models is often adjusted. For example, BAM~\cite{xin2020binarized} and BTM~\cite{jiang2021training} introduce multiple feature map aggregations and skip connection layers.

Except for 1-bit quantization, other ultra-low quantization precision such as 2-bit, 3-bit and 4-bit is discussed in many studies as well.
To handle the unstable activation ranges, Li~\cite{li2020pams} proposed a symmetric layer-wise linear quantizer that adopts a trainable clipping bound to clamp the abnormal activations. As for weights, the same symmetric quantizer is adopted but the clipping variable is simply set to the maximum magnitude of the weights. Moreover, the quantized model is enhanced by the structured knowledge transfer from its full-precision counterpart.
Wang~\emph{et al.}~\cite{wang2021fully} chose to quantize all layers of SR models and perform both weights and activations quantization using a symmetric layer-wise quantizer equipped with a trainable clipping variable.
DAQ~\cite{hong2022daq} observes that each channel has non-zero distributions and the activation values also vary drastically to the input image. Based on this observation, a channel-wise distribution-aware quantizer is adopted where the activations are normalized before discretizing and de-normalized after convolution.

\section{Methodology}

\subsection{Preliminaries}
\label{sec:Preliminaries}

Previous low-bit SR adopt a symmetric quantizer to perform quantization upon the premise of activations and weights with a symmetric distribution.
Specifically, denoting $b$ as the bit-width, $\bm{x}$ as weights or activations, $\alpha$ as the clipping bound, the symmetric linear quantizer is defined as:
\begin{equation}
\bar{\bm{x}} = round\big(\frac{clip(\bm{x}, \alpha)}{s}\big) \cdot s, 
\label{old quantizer}
\end{equation}
where $clip(\bm{x}, \alpha) = min\big(max(\bm{x}, -\alpha), \alpha\big)$, $\bar{\bm{x}}$ is the corresponding de-quantized value of $\bm{x}$, $round(\cdot)$ rounds its input to the nearest integer and $s$ denotes the scaling factor that projects a floating-point number to a fixed-point integer which can be calculated as $s = \frac{2\alpha}{2^{b-1}-1}$.

An appropriate clipping bound is crucial to the quantization performance since it not only eliminates outlier activations but refers to retaining details in the reconstructed images. To this end, PAMS~\cite{li2020pams} regards $\alpha$ as a learnable variable in quantizing activations, while setting $\alpha$ to the maximum magnitude of the weights in quantizing weights.

\subsection{Our Insights}
\label{sec:Our Insights}

Despite the progress, the performance of earlier low-bit SR quantization methods~\cite{li2020pams,hong2022daq,wang2021fully} remains an open issue when quantizing the full-precision counterpart to ultra-low precision with a layer-wise quantizer. After an in-depth analysis, we attribute the poor performance to the contradiction between the layer-wise symmetric quantizer and the asymmetric activation distribution in SR models.

Concretely, we observe the activations from different SR models. Fig.\,\ref{visualization} shows the histograms, maximum and minimum of activations collected from the pre-trained EDSR~\cite{lim2017enhanced} and RDN~\cite{zhang2018residual} on DIV2K~\cite{timofte2017ntire} benchmark. 
From the first column of Fig.\,\ref{visualization}, we realize that the activations of SR models are indeed highly asymmetric in an irregular state. For example, the magnitude of the minimum activation is almost twice that of the maximum activation in EDSR. A similar phenomenon can be observed in RDN.
Furthermore, as shown in the second and third columns of Fig.\,\ref{visualization}, the maximum and minimum of activations also vary drastically for different samples. 
Such an asymmetric activation distribution mostly results from the removal of BN layers in modern SR networks since BN reduces the scale information of images which however is crucial to SR tasks.
Though some studies retain BN layers, however, other remedies have to be taken to regain the performance. For example, SRResNet~\cite{ledig2017photo} propagates outputs of the first layer to the outputs of all following blocks, which indeed causes asymmetric distributions as well.
It is natural that the symmetric quantizer in existing studies cannot well fit the symmetric activation distributions, which we term as ``quantization unfitness'' in this paper.

To be specific, using only one clipping bound fails to handle the symmetric activations, whatever the value of $\alpha$.
To demonstrate this, we quantize the activations of RDN~\cite{zhang2018residual} to 3-bit using the quantizer in Eq.\,(\ref{old quantizer}). 
As displayed in Fig.\,\ref{problem:a}, the lack of activations along the negative axis direction causes a waste of two quantization levels if $\alpha$ is set to a large value such as the maximum of the activation magnitude.
%
%
The wastes are even more severe as the bit-width goes down. Taking the activations in Fig.\,\ref{problem:a} as an example, we experimentally observe $37.5\%$ are wasted in 3-bit quantization, while it increases sharply to $50\%$ in 2-bit quantization.
%
%
When it comes to a small $\alpha$ such as the absolute value of the minimum of activation magnitude, as illustrated in Fig.\,\ref{problem:b}, though avoiding the waste on quantization levels, only the small-magnitude activations are covered, leading to large quantization error since many high-magnitude activations are quantized to a small quantization level, \emph{i.e.}, $\alpha$.
%
%
%
Recall that SR models are sensitive to the scale information of images. Consequently, representing the large full-precision activations with a small quantization level inevitably brings about detail loss in the reconstructed high-resolution images, leading to significant quality degradation.
Similar observations can be found if the same clipping bound is applied to all the input images since the maximum and minimum activations over different images also drastically vary as illustrated in Fig.\,\ref{visualization}.

Overall, an appropriate quantizer is vital to the final performance of a quantized SR model.

\subsection{Our Solutions}

\begin{figure*}[t]
\centering
\subfigure[]{
\includegraphics[width=0.45\linewidth]{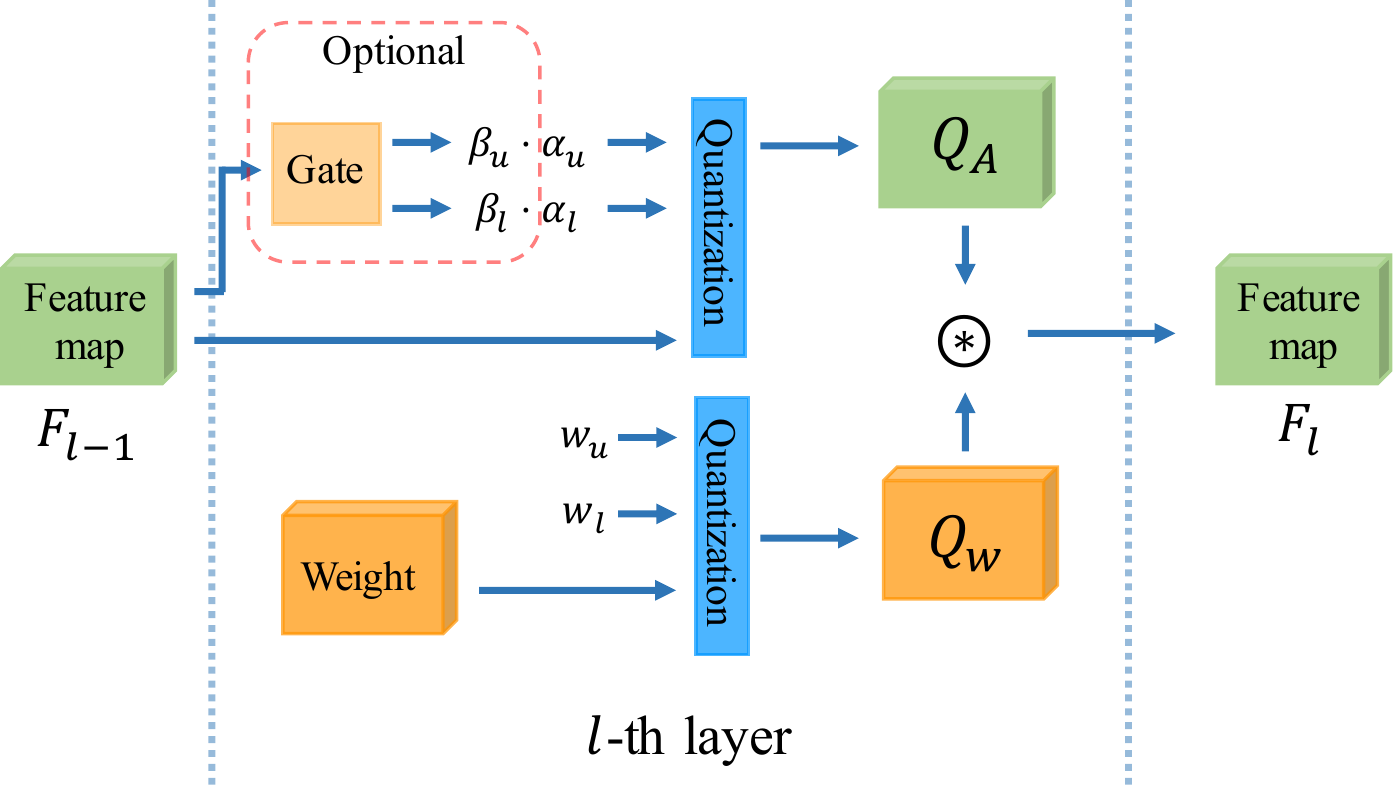} \label{framework:process}
}
\hspace{3mm}
\subfigure[]{
\includegraphics[width=0.45\linewidth]{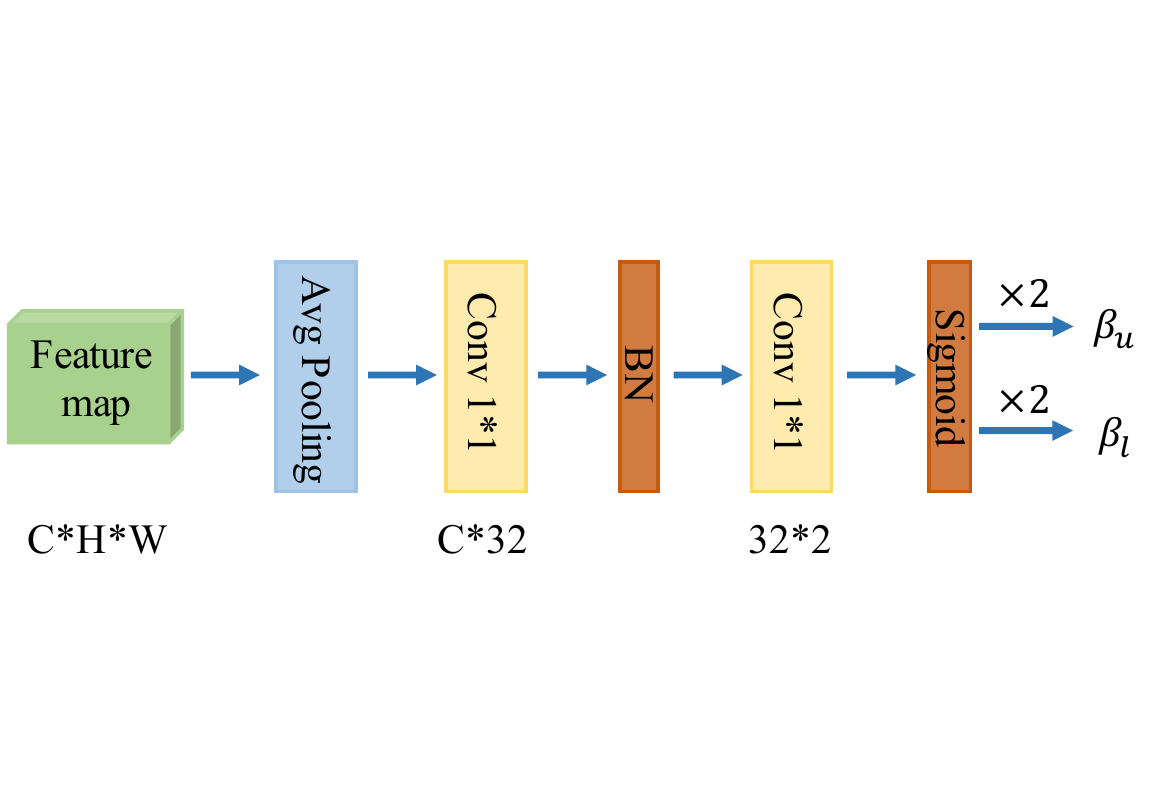}
\label{framework:gate} 
}
\vspace{-1.0em}
\caption{(a) Computational graph in the $l$-th quantized convolutional layer. (b) Structure of the dynamic gate controller. }
\label{framework}
\vspace{-1.5em}
\end{figure*}

In the following, we first detail dynamic dual trainable bounds (DDTB) specifically designed to quantize activations of SR models. Then, we elaborate on our initializer for DDTB. Finally, we describe the quantizer for weights. The overall computational graph is presented in Fig.\,\ref{framework:process}.

\textbf{Activation Quantization}.
\label{sec:Q of A}
Our DDTB consists of two parts: 1) A layer-wise quantizer with a trainable upper bound and a trainable lower bound; 2) A dynamic gate controller with adaptive upper and lower bounds to the inputs.

\textit{Dual trainable bounds:} As mentioned in Sec.\,\ref{sec:Our Insights}, only using one clipping variable to clamp the asymmetric activations leads to the ``quantization unfitness'' problem. To address this, we introduce two trainable clipping variables of $\alpha_u$ and $\alpha_l$ to respectively determine the upper bound and the lower bound of the activations. Equipped with these two dual trainable clipping variables, the quantized integer $\bm{q}$ for the input $\bm{x}$ can be obtained as:
\begin{equation}
\bm{q} = round\big(\frac{clip(\bm{x}, \alpha_l, \alpha_u)}{s}\big)+Z,
\label{new quantizer:q}
\end{equation}
where $clip(\bm{x}, \alpha_l, \alpha_u) = min\big(max(\bm{x}, \alpha_l), \alpha_u\big)$, $Z=round(\frac{-\alpha_l}{s})$ is a zero-point integer corresponding to the real value $0$. Accordingly, the scaling factor $s$ is calculated as: $s = \frac{\alpha_u - \alpha_l}{2^{b}-1}$. Finally, the de-quantized value $\bar{\bm{a}}$ is:
\begin{equation}
\bar{\bm{a}} = (\bm{q} - Z) \cdot s.
\label{new quantizer:deq}
\end{equation}

Combining Eq.\,(\ref{new quantizer:q}) and Eq.\,(\ref{new quantizer:deq}) completes our activation quantizer. 
As depicted in Fig.\,\ref{problem:dual}, the quantization levels generated by this quantization scheme well fit the activations when $\alpha_u$ and $\alpha_l$ are set to the maximum and minimum of activations, respectively. In order to accommodate samples with drastically varying ranges, we employ the stochastic gradient descent to adaptively learn $\alpha_u$ and $\alpha_l$ by minimizing the finally loss function. Denoting the clipped activations as $\Tilde{\bm{a}}=min(max(\bm{a}, \alpha_l), \alpha_u)$, $\frac{\partial \bar{\bm{a}}}{\partial \Tilde{\bm{a}}}$ is set to 1 by using the straight-through estimator (STE)~\cite{courbariaux2016binarized}. Then, the gradients of $\alpha_u$, $\alpha_l$ can be calculated as:
\begin{equation}
\begin{aligned}
    \frac{\partial \bar{\bm{a}}}{\partial \alpha_u} = \frac{\partial \bar{\bm{a}}}{\partial \Tilde{\bm{a}}} \frac{\partial \Tilde{\bm{a}}}{\partial \alpha_u} \overset{STE}{\approx} \begin{cases}
    1, & \bm{a} \geq \alpha_u\\
    0, & \bm{a} \leq \alpha_u
    \end{cases}, \quad
    \frac{\partial \bar{\bm{a}}}{\partial \alpha_l} = \frac{\partial \bar{\bm{a}}}{\partial \Tilde{\bm{a}}} \frac{\partial \Tilde{\bm{a}}}{\partial \alpha_l}
    \overset{STE}{\approx} \begin{cases}
    0, & \bm{a} \geq \alpha_l\\
    1, & \bm{a} \leq \alpha_l
    \end{cases}.
\end{aligned}
\label{gd_ori}
\end{equation}

With the dual $\alpha_l$ and $\alpha_u$, ``quantization unfitness'' can be alleviated.
Note that, early study~\cite{jung2019learning} also introduces two trainable parameters to determine the quantization range. This paper differs in: 
1) \cite{jung2019learning} utilizes the center and radius of the quantization region to parameterize the quantization range while we explicitly introduce the upper and lower bounds. 2) \cite{jung2019learning} requires an expensive transformation to normalize activations to a fixed range before performing quantization. 3) Though trainable, the final quantization range is consistent with different inputs. Thus, we design a dynamic gate controller to provide an instance-wise quantization range adaptive to different inputs as illustrated in the next section.

\textit{Dynamic gate controller:} 
Though the trainable upper and lower bounds partially alleviate the ``quantization unfitness'' problem, it is still suboptimal if only applying the same pair of ($\alpha_l$, $\alpha_u$) for different inputs since their activation ranges also dramatically vary as shown in Fig.\,\ref{visualization}.

To well address this issue, we devise a dynamic gate controller to adapt ($\alpha_l$, $\alpha_u$) to every input sample at runtime. Fig.\,\ref{framework:gate} depicts the structure of our gate controller, which is composed of a series of consequently stacked operators including two convolutional layers, a BN layer, a ReLU function, and a sigmoid function multiplied by 2.
It takes the feature maps from the SR model as inputs and outputs two scaling coefficients $\beta_l$ and $\beta_u$ for each input image. 
Then,  $\beta_l$ and $\beta_u$ are respectively used to re-scale the lower bound $\alpha_l$ and the upper bound $\alpha_u$, results of which serve as the final clipping bounds of the corresponding input images as shown in Fig.\,\ref{framework:process}.
%
%
Denoting the $\alpha_u' = \beta_u\cdot\alpha_u$ and $\alpha_l' = \beta_l\cdot\alpha_l$, the clipped activations $\Tilde{\bm{a}}$ now can be reformulated as: $\Tilde{\bm{a}} = min(max(\bm{a}, \alpha_l'), \alpha_u')$. Similar to Eq.\,(\ref{gd_ori}), the gradient can be calculated as:
\begin{equation}
\begin{aligned}
    \frac{\partial \bar{\bm{a}}}{\partial \alpha_u'} = \frac{\partial \bar{\bm{a}}}{\partial \Tilde{\bm{a}}} \frac{\partial \Tilde{\bm{a}}}{\partial \alpha_u'} \overset{STE}{\approx} \begin{cases}
    1, & \bm{a} \geq \alpha_u'\\
    0, & \bm{a} \leq \alpha_u'
    \end{cases}, \quad
    \frac{\partial \bar{\bm{a}}}{\partial \alpha_l'} = \frac{\partial \bar{\bm{a}}}{\partial \Tilde{\bm{a}}} \frac{\partial \Tilde{\bm{a}}}{\partial \alpha_l'}
    \overset{STE}{\approx} \begin{cases}
    0, & \bm{a} \geq \alpha_l'\\
    1, & \bm{a} \leq \alpha_l'
    \end{cases}.
\end{aligned}
\label{gd_rescale}
\end{equation}

Then, the gradients of $\beta_u$, $\beta_l$, $\alpha_u$, $\alpha_l$ can be obtained by the chain rule. With these gradients, the gate and clipping bounds can be trained together.

Our gate controller enables the clipping bounds adaptive to each of the input samples, with which the gate outputs are dynamically correlated at runtime. 
Nevertheless, the extra costs on computation and storage from the introduced gate controller are expensive, causing a contradiction with our motive to reduce the complexity of SR models. 
Consequently, we reduce the gate complexity from two perspectives:
1) We quantize the weights and activations of the gate to a 2-bit using the quantizer defined by Eq.\,(\ref{new quantizer:q}) and Eq.\,(\ref{new quantizer:deq}). The clipping bounds are simply set as the maximum and minimum of the weights/activations. We empirically find that such a simple setting is sufficient. During the inference phase, the BN layer in our gate controller can be folded into the previous convolutional layer~\cite{wang2021fully}. Therefore, the extra floating-point operations are very cheap in this situation.
2) We choose to apply our dynamic gate controller to some layers of the SR model. Concretely, we define the dynamic intensity in the $l$-th layer as $DI^l = V^l_{max} + V^l_{min}$, where $V^l_{max}$ and $V^l_{min}$ are variances of the maximum activations and minimum activations.
%
%
The overhead is trivial as it forwards the training data to the full-precision network only once and can be completed offline. It is intuitive that a larger $DI^l$ indicates a more dynamic activation range. Thus, we apply the dynamic gate to these layers with the top-$P\%$ largest dynamic intensity, where the gate ratio $P$ is a hyper-parameter.

\textbf{DDTB Initializer.} 
One common way to initialize the $\alpha_u$ and $\alpha_l$ is to use the activation statistics such as the maximum and minimum values in the quantized network. However, for ultra-low bit cases, quantization error accumulates drastically, leading to unreliable statistics in deep layers. 
Instead, we derive the activation statistics by feeding a patch of images to the full-precision network, and then use the $M$-th and $(100-M)$-th percentiles of obtained activations to initialize the $\alpha_u$ and $\alpha_l$, which are more stable even in the ultra-low bit situation since the full-precision network is not challenged by the quantization error issue.

In addition, we observe that the initial $\beta_u$ and $\beta_l$ are either too small or too large in the early training phase due to the random initialization of gate weights, which obstacles the learning of $\alpha_u$ and $\alpha_l$, and further leads to inferior performance. 
Therefore, in the first $K$ training epochs, we do not apply $\beta_u$ and $\beta_l$ to scaling $\alpha_u$, $\alpha_l$, and the gate is trained individually to push its outputs close to 1 whatever the inputs. After $K$ epochs, we then apply $\beta_u$ and $\beta_l$ which start from a stable initial value, \emph{i.e.}, around 1. Then, the dynamic gate controller and the clipping variables are jointly trained to accommodate the drastically varying activation ranges. 
In all our experiments, we set $K = 5$.

\textbf{Weight Quantization}.
\label{sec:Q of W}
The symmetric quantizer is also adopted to quantize weights in previous works. We suggest utilizing the asymmetric quantizer to quantize the weights. Similar to our activation quantization, an upper bound $w_u$ and a lower bound $w_l$ are used to clip the weight range: $\Tilde{\bm{w}} = min(max(\bm{w}, w_l), w_u)$. The quantized integer and de-quantized value can be obtained similar to Eq.\,(\ref{new quantizer:q}) and Eq.\,(\ref{new quantizer:deq}).
This quantization scheme is also compatible with the symmetric case when $w_u=-w_l$.
In all our experiments, $w_u$ is set to the $99$-th percentile and $w_u$ is set to $1$-th percentile of the full-precision weights since we observe no performance gains if setting them to trainable parameters.

\subsection{Training Loss}
Following PAMS~\cite{li2020pams}, we use $L_1$ loss and $L_{SKT}$ loss in all the experiments. Denoting $D=\{I^i_{LR}, I^i_{HR}\}_{i=1}^N$ as the training dataset with $N$ pairs of LR and HR images. The $L_1$ loss and $L_{SKT}$ loss are defined as:
\begin{equation}
L_1 = \frac{1}{N}\sum_{i=1}^N\| I^i_{LR} - I^i_{HR} \|_1,
\label{loss:l1}
\end{equation}
\begin{equation}
L_{SKT} = \frac{1}{N}\sum_{i=1}^N\| \frac{F_S'(I^i_{LR})}{\|F_S'(I^i_{LR})\|_2} - \frac{F_T'(I^i_{LR})}{\|F_T'(I^i_{LR})\|_2} \|_2,
\label{loss:SKT}
\end{equation}
where $F_S'(I^i_{LR})$ and $F_T'(I^i_{LR})$ are structure features of $I^i_{LR}$ from the quantized network and the full-precision network, respectively. The structure feature can be calculated by $F'(I^i_{LR})=\sum_{c=1}^N|F_c(I^i_{LR})|^2  \in \mathbb R^{H \times W}$, where $F(I^i_{LR}) \in \mathbb R^{C \times H \times W}$ is the feature map after the last layer in the high-level feature
extractor. Then, the overall loss function $L$ is:
\begin{equation}
L = L_1 + \lambda L_{SKT}.
\label{loss:overall}
\end{equation}
where $\lambda = 1,000$ in all our experiments.

\begin{table}[!t]
\setlength\tabcolsep{3.0pt}
\scriptsize
\centering
\caption{PSNR/SSIM comparisons between existing low-bit SR methods and our DDTB in quantizing EDSR~\cite{lim2017enhanced} of scale 4 and scale 2 to the low-bit format. Results of the full-precision model are displayed below the dataset name.}
\vspace{-1.0em}
\begin{tabular}{cccccccc}
\toprule[1.25pt]
Model                                                               & Dataset                   &Bit & Dorefa~\cite{zhou2016dorefa}        & TF Lite~\cite{jacob2018quantization}       & PACT~\cite{choi2018pact}          & PAMS~\cite{li2020pams}          & \textbf{DDTB(Ours)}          \\ \hline \hline
\multirow{12}{*}{\begin{tabular}[c]{@{}c@{}}EDSR\\ $\times$4\end{tabular}} & \multirow{3}{*}{\begin{tabular}[c]{@{}c@{}}Set5~\cite{bevilacqua2012low}\\ 32.10/0.894\end{tabular}}     & 2         & 29.90/0.850 & 29.96/0.851 & 30.03/0.854 & 29.51/0.835 & \textbf{30.97/0.876} \\
                                                                    &                           & 3         & 30.76/0.870        & 31.05/0.877        & 30.98/0.876        &  27.25/0.780        & \textbf{31.52/0.883}        \\
                                                                    &                           & 4         & 30.91/0.873        & 31.54/0.884        & 31.32/0.882        & 31.59/0.885        & \textbf{31.85/0.889}        \\ \cline{2-8} 
                                                                    & \multirow{3}{*}{\begin{tabular}[c]{@{}c@{}}Set14~\cite{ledig2017photo}\\ 28.58/0.781\end{tabular}}    & 2         & 27.08/0.744        & 27.12/0.745        &  27.21/0.747        & 26.79/0.734        & \textbf{27.87/0.764}        \\
                                                                    &                           & 3         & 27.66/0.759        & 27.92/0.765        & 27.87/0.764        & 25.24/0.673        & \textbf{28.18/0.771}        \\
                                                                    &                           & 4         & 27.78/0.762        & 28.20/0.772        & 28.07/0.769        & 28.20/0.773        & \textbf{28.39/0.777}        \\ \cline{2-8} 
                                                                    & \multirow{3}{*}{\begin{tabular}[c]{@{}c@{}}BSD100~\cite{martin2001database}\\ 27.56/0.736\end{tabular}}   & 2         & 26.66/0.704        &  26.68/0.705        & 26.73/0.706        & 26.45/0.696        & \textbf{27.09/0.719}        \\
                                                                    &                           & 3         & 26.97/0.716        & 27.12/0.721        & 27.09/0.719        & 25.38/0.644        & \textbf{27.30/0.727}        \\
                                                                    &                           & 4         & 27.04/0.719        &  27.31/0.727        & 27.21/0.724        & 27.32/0.728        & \textbf{27.44/0.732}        \\ \cline{2-8} 
                                                                    & \multirow{3}{*}{\begin{tabular}[c]{@{}c@{}}Urban100~\cite{huang2015single}\\ 26.04/0.785\end{tabular}} &  2         & 24.02/0.705        & 24.03/0.705        & 24.12/0.708        & 23.72/0.688        & \textbf{24.82/0.742}        \\
                                                                    &                           & 3         & 24.59/0.732        & 24.85/0.743        & 24.82/0.741        & 22.76/0.641        &  \textbf{25.33/0.761}      \\
                                                                    &                           & 4         & 24.73/0.739        & 25.28/0.760        & 25.05/0.751        & 25.32/0.762        & \textbf{25.69/0.774}        \\ \hline \hline
\multirow{12}{*}{\begin{tabular}[c]{@{}c@{}}EDSR\\ $\times$2\end{tabular}} & \multirow{3}{*}{\begin{tabular}[c]{@{}c@{}}Set5~\cite{bevilacqua2012low}\\ 37.93/0.960\end{tabular}}     & 2         & 36.12/0.952 & 36.23/0.952 & 36.58/0.955 & 35.30/0.946 & \textbf{37.25/0.958} \\
                                                                    &                           & 3         & 37.13/0.957        & 37.33/0.957        & 37.36/0.958        & 36.76/0.955        & \textbf{37.51/0.958}        \\
                                                                    &                           & 4         & 37.22/0.958        & 37.64/0.959        & 37.57/0.958        & 37.67/0.958        & \textbf{37.72/0.959}        \\ \cline{2-8} 
                                                                    & \multirow{3}{*}{\begin{tabular}[c]{@{}c@{}}Set14~\cite{ledig2017photo}\\ 33.46/0.916\end{tabular}}    & 2         & 32.09/0.904        & 32.14/0.904        & 32.38/0.907        & 31.63/0.899        & \textbf{32.87/0.911}        \\
                                                                    &                           & 3         & 32.73/0.910        & 32.98/0.912        & 32.99/0.912        & 32.50/0.907        & \textbf{33.17/0.914}        \\
                                                                    &                           & 4         & 32.82/0.911        & 33.24/0.914        & 33.20/0.914        & 33.20/0.915        & \textbf{33.35/0.916}        \\ \cline{2-8} 
                                                                    & \multirow{3}{*}{\begin{tabular}[c]{@{}c@{}}BSD100~\cite{martin2001database}\\ 32.10/0.899\end{tabular}}   & 2         & 31.03/0.884        & 31.07/0.885        & 31.26/0.887        & 30.66/0.879        & \textbf{31.67/0.893}        \\
                                                                    &                           & 3         & 31.57/0.892        & 31.76/0.894        & 31.77/0.894        & 31.38/0.889      & \textbf{31.89/0.896}        \\
                                                                    &                           & 4         & 31.63/0.893        & 31.94/0.896        & 31.93/0.897        & 31.94/0.897        & \textbf{32.01/0.898}        \\ \cline{2-8} 
                                                                    & \multirow{3}{*}{\begin{tabular}[c]{@{}c@{}}Urban100~\cite{huang2015single}\\ 31.71/0.925\end{tabular}} & 2         & 28.71/0.886        & 28.77/0.886        & 29.22/0.894        & 28.11/0.875        & \textbf{30.34/0.910}        \\
                                                                    &                           & 3         & 30.00/0.906        & 30.48/0.912        & 30.57/0.912        &  29.50/0.898        & \textbf{31.01/0.919}        \\
                                                                    &                           & 4         & 30.17/0.908        & 31.11/0.919        & 31.09/0.919        & 31.10/0.919        & \textbf{31.39/0.922}        \\ \bottomrule[0.75pt]
\end{tabular}
\label{EDSR}
\end{table}

\section{Experiments}

\subsection{Implementation Details}

%

All the models are trained on the training set of DIV2K including 800 images~\cite{timofte2017ntire}, and tested on four standard benchmarks including Set5~\cite{bevilacqua2012low}, Set14~\cite{ledig2017photo}, BSD100~\cite{martin2001database} and Urban100~\cite{huang2015single}. Two upscaling factors of $\times$2 and $\times$4 are evaluated.
The quantized SR models include EDSR~\cite{lim2017enhanced}, RDN~\cite{zhang2018residual}, and SRResNet~\cite{ledig2017photo}. We quantize them to 4, 3, and 2-bit and compare with the SOTA competitors including DoReFa~\cite{zhou2016dorefa}, Tensorflow Lite (TF Lite)~\cite{jacob2018quantization}, PACT~\cite{choi2018pact}, and PAMS~\cite{li2020pams}.
The PSNR and SSIM~\cite{wang2004image} over the Y channel are reported. 

\begin{table}[!t]
\setlength\tabcolsep{3.0pt}
\scriptsize
\centering
\caption{PSNR/SSIM comparisons between existing low-bit SR methods and our DDTB in quantizing RDN~\cite{zhang2018residual} of scale 4 and scale 2 to the low-bit format. Results of the full-precision model are displayed below the dataset name.}
\vspace{-1.0em}
\begin{tabular}{cccccccc}
\toprule[1.25pt]
Model                                               & Dataset  &Bit & Dorefa~\cite{zhou2016dorefa}        & TF Lite~\cite{jacob2018quantization}       & PACT~\cite{choi2018pact}          & PAMS~\cite{li2020pams}         &\textbf{DDTB(Ours)}          \\ \hline \hline
\multirow{12}{*}{\begin{tabular}[c]{@{}c@{}}RDN\\ $\times$4\end{tabular}} & \multirow{3}{*}{\begin{tabular}[c]{@{}c@{}}Set5~\cite{bevilacqua2012low}\\ 32.24/0.896\end{tabular}}     & 2         & 29.90/0.849 & 29.93/0.850 & 28.78/0.820 & 29.73/0.843 & \textbf{30.57/0.867} \\
  &                           & 3         & 31.24/0.881        & 30.13/0.854        & 31.30/0.879        &  29.54/0.838        & \textbf{31.49/0.883}        \\
                                                                    &                           & 4         & 31.51/0.885        & 31.08/0.874        & 31.93/0.890        & 30.44/0.862        & \textbf{31.97/0.891}        \\ \cline{2-8} 
                                                                    & \multirow{3}{*}{\begin{tabular}[c]{@{}c@{}}Set14~\cite{ledig2017photo}\\ 28.67/0.784\end{tabular}}    & 2         & 27.08/0.743        & 27.11/0.744        & 26.33/0.717        & 26.96/0.739        & \textbf{27.56/0.757}        \\
                                                                    &                           & 3         & 28.02/0.769        & 27.21/0.744        & 28.06/0.767        & 26.82/0.734        & \textbf{28.17/0.772}        \\
                                                                    &                           & 4         & 28.21/0.773        & 27.98/0.764        & 28.44/0.778        & 27.54/0.753        & \textbf{28.49/0.780}        \\ \cline{2-8} 
                                                                    & \multirow{3}{*}{\begin{tabular}[c]{@{}c@{}}BDS100~\cite{martin2001database}\\ 27.63/0.738\end{tabular}}   & 2         & 26.65/0.703        & 26.64/0.703        & 26.16/0.681        & 26.57/0.700        & \textbf{26.91/0.714}        \\
                                                                    &                           & 3         & 27.20/0.724        & 26.71/0.705        & 27.21/0.722        & 26.47/0.696        & \textbf{27.30/0.728}        \\
                                                                    &                           & 4         & 27.30/0.727        & 27.16/0.720        & 27.46/0.732        & 26.87/0.710        & \textbf{27.49/0.735}        \\ \cline{2-8} 
                                                                    & \multirow{3}{*}{\begin{tabular}[c]{@{}c@{}}Urban100~\cite{huang2015single}\\ 26.29/0.792\end{tabular}} & 2         & 23.99/0.702        & 23.99/0.703        & 23.38/0.672        & 23.87/0.696        & \textbf{24.50/0.728}        \\
                                                                    &                           & 3         & 25.07/0.754        & 24.27/0.713        & 25.17/0.754        & 23.83/0.692        & \textbf{25.35/0.764}        \\
                                                                    &                           & 4         & 25.36/0.764        & 25.25/0.755        & 25.83/0.779        & 24.52/0.726        & \textbf{25.90/0.783}        \\ \hline \hline
\multirow{12}{*}{\begin{tabular}[c]{@{}c@{}}RDN\\ $\times$2\end{tabular}} & \multirow{3}{*}{\begin{tabular}[c]{@{}c@{}}Set5~\cite{bevilacqua2012low}\\ 38.05/0.961\end{tabular}}     & 2         & 36.20/0.952 & 36.12/0.951 & 36.55/0.954 & 35.45/0.946 & \textbf{36.76/0.955} \\
                                                                    &                           & 3         & 37.44/0.958        & 36.38/0.953        & 37.39/0.958        & 35.25/0.942        & \textbf{37.61/0.959}        \\
                                                                    &                           & 4         & 37.61/0.959        & 36.83/0.955        & 37.82/0.959        & 36.53/0.953        & \textbf{37.88/0.960}        \\ \cline{2-8} 
                                                                    & \multirow{3}{*}{\begin{tabular}[c]{@{}c@{}}Set14~\cite{ledig2017photo}\\ 33.59/0.917\end{tabular}}    & 2         & 32.14/0.904        & 32.08/0.903        & 32.34/0.905        & 31.67/0.899        & \textbf{32.54/0.908}        \\
                                                                    &                           & 3         & 33.08/0.914        & 32.46/0.907        & 33.08/0.914        & 31.52/0.893        & \textbf{33.26/0.915}        \\
                                                                    &                           & 4         & 33.23/0.915        & 32.72/0.910        & 33.47/0.916        & 32.39/0.905        & \textbf{33.51/0.917}        \\ \cline{2-8} 
                                                                    & \multirow{3}{*}{\begin{tabular}[c]{@{}c@{}}BSD100~\cite{martin2001database}\\ 32.20/0.900\end{tabular}}   & 2         & 31.06/0.885        & 31.01/0.884        & 31.21/0.886        & 30.69/0.879        & \textbf{31.44/0.890}        \\
                                                                    &                           & 3         & 31.87/0.896        & 31.34/0.888        & 31.86/0.896        & 30.62/0.874        & \textbf{31.91/0.897}        \\
                                                                    &                           & 4         & 31.98/0.897        & 31.63/0.892        & 32.09/0.898        & 31.27/0.885        & \textbf{32.12/0.899}        \\ \cline{2-8} 
                                                                    & \multirow{3}{*}{\begin{tabular}[c]{@{}c@{}}Urban100~\cite{huang2015single}\\ 32.13/0.927\end{tabular}} & 2         & 28.81/0.888        & 28.72/0.885        & 29.15/0.892        & 28.14/0.874        & \textbf{29.77/0.903}        \\
                                                                    &                           & 3         & 30.96/0.918        & 29.83/0.903        & 30.97/0.918        & 28.30/0.873        & \textbf{31.10/0.920}        \\
                                                                    &                           & 4         & 31.33/0.922        & 30.49/0.912        & 31.69/0.925        & 29.70/0.898        & \textbf{31.76/0.926}        \\ \bottomrule[0.75pt]
\end{tabular}
\label{RDN}
\end{table}

The full-precision models and compared quantization methods are implemented based on their open-source code. For the quantized models, following PAMS~\cite{li2020pams}, we quantize both weights and activations of the high-level feature extraction module.
The low-level feature extraction and reconstruction modules are set to the full-precision\footnote{Results of fully quantized SR models are provided in the Appendix.}. The batch size is set to 16 and the optimizer is Adam~\cite{kingma2014adam} with $\beta_1=0.9$, $\beta_2=0.999$ and $\epsilon=10^{-8}$. We set the initial learning rate to $10^{-4}$ and halve it every 10 epochs. For EDSR, the gate ratio $P$ is set to $30$ and the initialization coefficient $M$ is $99$. As for RDN, $P$ and $M$ are $50$ and $95$. For SRResNet, $P=10$ and $M=99$. The total training epochs are set to 60. The training images are pre-processed by subtracting the mean RGB. During training, random horizontal flip and vertical rotation are adopted to augment data. All experiments are implemented with PyTorch~\cite{paszke2019pytorch}.

\subsection{Experimental Results}
\label{sec:experimental results}
Table\,\ref{EDSR}, Table\,\ref{RDN}, and Table\,\ref{SRResNet} respectively show the quantitative results of EDSR, RDN, and SRResNet on different datasets. As can be seen, our DDTB consistently outperforms all the compared methods on these quantized SR models with different bit-widths. More details are discussed below.

\textbf{Evaluation on EDSR}.
In the case of 4-bit, our DDTB outperforms the advanced PAMS by a large margin. For instance, for 4-bit EDSR$\times$4, DDTB obtains 0.37dB PSNR gains on Urban100. More noticeable improvements can be observed when performing ultra-low bit quantization. For instance, our DDTB obtains performance gains by 0.94dB, 0.66dB, 0.36dB, and 0.70dB on Set5, Set14, BSD100, and Urban100 when quantizing EDSR$\times$4 to 2-bit.

\textbf{Evaluation on RDN}.
When quantizing the model to 4-bit, our DDTB slightly outperforms the existing SOTA of PACT. When it comes to ultra-low bit, the superior performance is in particular obvious. In detail, for 2-bit RDN$\times$4, the performance gains of our DDTB are 0.64dB, 0.45dB, 0.26dB, and 0.51dB on Set5, Set14, BSD100, and Urban100.

\textbf{Evaluation on SRResNet}.
The results of SRResNet also manifest that the performance gains of our DDTB are more prominent with ultra-low precision. For 2-bit SRResNet$\times$4, our DDTB improves the performance by 0.65dB, 0.47dB, 0.30dB, and 0.69dB on Set5, Set14, BSD100, and Urban100, while the performance gains are 1.15dB, 0.79dB, 0.67dB, and 1.80dB for 2-bit SRResNet$\times$2.

\begin{table}[!t]
\setlength\tabcolsep{2.1pt}
\scriptsize
\centering
\caption{PSNR/SSIM comparison between existing low-bit SR methods and our DDTB in quantizing SRResNet~\cite{ledig2017photo} of scale 4 and scale 2 to the low-bit format. Results of the full-precision model are displayed below the dataset name.}
\vspace{-1.0em}
\begin{tabular}{cccccccc}
\toprule[1.25pt]
Model                                                               & Dataset                   & Bit & Dorefa~\cite{zhou2016dorefa}        & TF Lite~\cite{jacob2018quantization}       & PACT~\cite{choi2018pact}          & PAMS\cite{li2020pams}          & \textbf{DDTB(Ours)}         \\ \hline \hline
\multirow{12}{*}{\begin{tabular}[c]{@{}c@{}}SRResNet\\$\times$4\end{tabular}} & \multirow{3}{*}{\begin{tabular}[c]{@{}c@{}}Set5~\cite{bevilacqua2012low}\\ 32.07/0.893\end{tabular}}     & 2         & 30.25/0.860 & 30.33/0.861 & 30.86/0.874 & 30.25/0.861 & \textbf{31.51/0.887} \\
                                                                    &                           & 3         & 30.34/0.862        & 31.58/0.886        & 31.62/0.887        & 31.68/0.888        & \textbf{31.85/0.890}        \\
                                                                    &                           & 4         & 30.32/0.861        & 31.82/0.890        & 31.85/0.890        & 31.88/0.891        & \textbf{31.97/0.892}        \\ \cline{2-8} 
                                                                    & \multirow{3}{*}{\begin{tabular}[c]{@{}c@{}}Set14~\cite{ledig2017photo}\\ 28.50/0.780\end{tabular}}    & 2         & 27.33/0.749        & 27.39/0.751        & 27.76/0.761        & 27.36/0.750        & \textbf{28.23/0.773}        \\
                                                                    &                           & 3         & 27.39/0.751        & 28.24/0.773        &  28.25/0.773        & 28.27/0.774        & \textbf{28.39/0.776}        \\
                                                                    &                           & 4         & 27.41/0.751        & 28.40/0.777        & 28.38/0.775        & 28.41/0.777        & \textbf{28.46/0.778}        \\ \cline{2-8} 
                                                                    & \multirow{3}{*}{\begin{tabular}[c]{@{}c@{}}BSD100~\cite{martin2001database}\\ 27.52/0.735\end{tabular}}   & 2         & 26.78/0.708        & 26.80/0.709        & 27.03/0.717        & 26.79/0.709        & \textbf{27.33/0.728}        \\
                                                                    &                           & 3         & 26.81/0.709        & 27.31/0.726        & 27.33/0.727        & 27.34/0.728        & \textbf{27.44/0.731}        \\
                                                                    &                           & 4         & 26.82/0.709        & 27.42/0.730        & 27.41/0.730        & 27.45/0.732        & \textbf{27.48/0.733}        \\ \cline{2-8} 
                                                                    & \multirow{3}{*}{\begin{tabular}[c]{@{}c@{}}Urban100~\cite{huang2015single}\\ 25.86/0.779\end{tabular}} & 2         & 24.17/0.711        & 24.21/0.713        & 24.68/0.733        & 24.19/0.713        & \textbf{25.37/0.762}        \\
                                                                    &                           & 3         & 24.24/0.714        & 25.33/0.759        & 25.39/0.761        & 25.46/0.765        & \textbf{25.64/0.770}        \\
                                                                    &                           & 4         & 24.26/0.714        & 25.62/0.780        & 25.61/0.769        & 25.68/0.773        & \textbf{25.77/0.776}        \\ \hline \hline
\multirow{12}{*}{\begin{tabular}[c]{@{}c@{}}SRResNet\\$\times$2\end{tabular}} & \multirow{3}{*}{\begin{tabular}[c]{@{}c@{}}Set5~\cite{bevilacqua2012low}\\ 37.89/0.960\end{tabular}}     & 2         & 35.27/0.946 & 35.34/0.946 & 36.31/0.953 & 34.75/0.942 & \textbf{37.46/0.958} \\
                                                                    &                           & 3         & 35.30/0.946        & 37.50/0.958        & 37.42/0.958        & 37.52/0.958        & \textbf{37.67/0.959}        \\
                                                                    &                           & 4         & 35.39/0.946        & 37.69/0.959        & 37.65/0.959        & 37.71/0.959        & \textbf{37.78/0.960}        \\ \cline{2-8} 
                                                                    & \multirow{3}{*}{\begin{tabular}[c]{@{}c@{}}Set14~\cite{ledig2017photo}\\ 33.40/0.916\end{tabular}}    & 2         & 31.54/0.899        & 31.61/0.899        & 32.23/0.905        & 31.31/0.896        & \textbf{33.02/0.913}        \\
                                                                    &                           & 3         & 31.56/0.899        & 33.05/0.913        & 32.92/0.911        & 33.09/0.914        & \textbf{33.24/0.915}        \\
                                                                    &                           & 4         & 31.63/0.899        & 33.26/0.915        & 33.24/0.915        & 33.26/0.915        & \textbf{33.32/0.916}        \\ \cline{2-8} 
                                                                    & \multirow{3}{*}{\begin{tabular}[c]{@{}c@{}}BSD100~\cite{martin2001database}\\ 32.08/0.898\end{tabular}}   & 2         & 30.61/0.879        & 30.66/0.879       & 31.11/0.885        & 30.48/0.877        & \textbf{31.78/0.895}        \\
                                                                    &                           & 3         & 30.62/0.879        & 31.81/0.894        & 31.70/0.893        & 31.85/0.896        & \textbf{31.95/0.897}        \\
                                                                    &                           & 4         & 30.67/0.880        & 31.99/0.897        & 31.96/0.897        & 31.99/0.897        & \textbf{32.03/0.898}        \\ \cline{2-8} 
                                                                    & \multirow{3}{*}{\begin{tabular}[c]{@{}c@{}}Urban100~\cite{huang2015single}\\ 31.60/0.923\end{tabular}} & 2         & 27.98/0.871        & 28.04/0.872        & 28.77/0.885        & 27.86/0.868        & \textbf{30.57/0.913}        \\
                                                                    &                           & 3         & 27.99/0.871        & 30.64/0.913        & 30.43/0.910        & 30.69/0.914        & \textbf{31.15/0.919}        \\
                                                                    &                           & 4         & 28.06/0.872        & 31.25/0.920        & 31.19/0.919        & 31.20/0.920        & \textbf{31.40/0.921}        \\ \bottomrule[0.75pt]
\end{tabular}
\label{SRResNet}
\end{table}

\textbf{Analysis}.
These results demonstrate in Table\,\ref{EDSR}, Table\,\ref{RDN}, and Table\,\ref{SRResNet} well demonstrate the effectiveness of our DDTB and also verify the correctness of our motivation in designing an appropriate quantizer adaptive to the activation distribution.
Moreover, it is worth noticing that DDTB provides more stable improvements over different SR models and bit-widths. 
Taking EDSR$\times$4 as an example, though PAMS obtains the best results among the compared methods when performing 4-bit quantization, it was outperformed by TF Lite in 3-bit and PACT in 2-bit. 
In contrast, DDTB achieves the best results in all these three bit-widths. Such stability further illustrates the advanced generalization of our DDTB.

\begin{figure*}[t]
\centering
\includegraphics[width=0.9\linewidth]{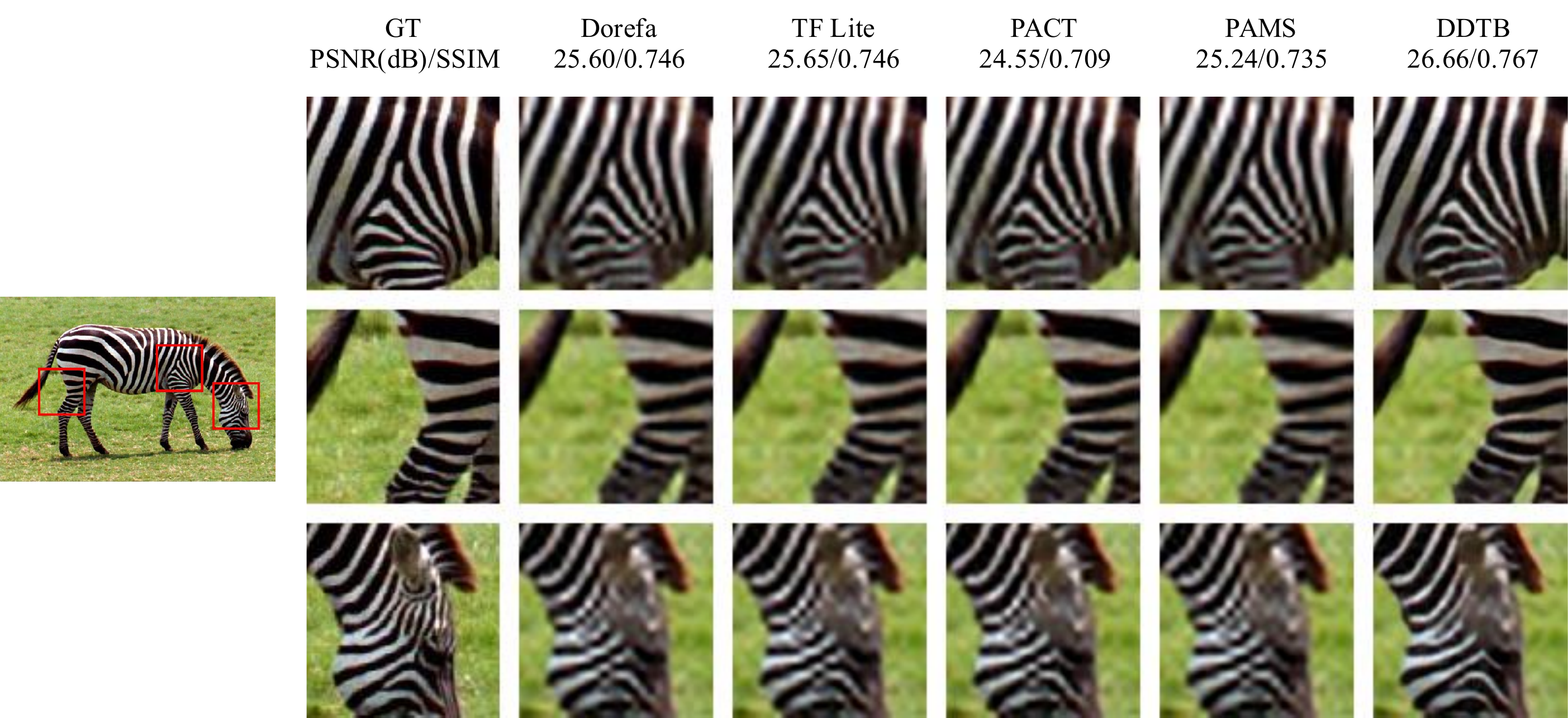}
\caption{Reconstructed results of 2-bit EDSR$\times$4}
\label{vis-edsr}
\end{figure*}

\textbf{Qualitative Visualizations}. Fig.\,\ref{vis-edsr} exhibits the qualitative visualizations of 2-bit EDSR\footnote{More qualitative visualizations are presented in the Appendix.}. Compared with others methods, the reconstructed HR image of our DDTB provides sharper edges and richer details.

\subsection{Model Analysis}

\textbf{Model Complexity}. To measure the complexity of the quantized network, we use the parameters and Bit-Operations (BOPs)~\cite{van2020bayesian} as the metric. BOPs are the number of multiplication operations multiplied by the bit-widths of two operands. As shown in Table\,\ref{analysis}, our DDTB significantly reduces the model size and computation cost. When being quantized to 2-bit, the size and computation costs of RDN are reduced by 92\% and 94\%, respectively. 
Moreover, the extra computation cost from our dynamic gate controller is negligible, as well as the gate size. For EDSR and SRResNet, the gate occupies less than 1\%  in model size, while for RDN, the gate only occupies 2.8\%.
Note that, the full-precision low-level feature extraction and reconstruction modules occupy most of the memory and computation costs when quantizing the network to the case of ultra-low bit.

\begin{table}[!t]
\setlength\tabcolsep{5pt}
\scriptsize
\centering
\caption{Complexity analysis. The number in brackets indicates the parameters in the high-level feature extraction module. We compute BOPs by generating a 1920$\times$1080 image (upscaling factor $\times$4). Results of 2-bit are displayed and more can be found in the Appendix.}
\vspace{-1.0em}
\begin{tabular}{ccccccc}
\toprule[1.25pt]
Model & Bit & Params & Gate Params($ratio$) & BOPs & Gate BOPs($ratio$)  \\  \hline \hline
EDSR~\cite{lim2017enhanced}  & 32  &    1.52M   &   0      &  532T   & 0          \\
EDSR\_DDTB & 2  &  0.41M(0.08M)    &   0.6\%  & 219T       &  0.0000013\%  \\ \hline
RDN~\cite{zhang2018residual}  & 32  &    22.3M  &   0      & 6038T  &    0      \\
RDN\_DDTB & 2  &  1.76M(1.42M)       &    2.8\%         &  239T &  0.0000066\%  \\ \hline
SRResNet~\cite{ledig2017photo}  & 32  &    1.543M  &   0      & 591T  &    0       \\
SRResNet\_DDTB & 2  & 0.44M(0.07M)     &   0.1\%    & 278T  &  0.0000002\%    \\
\bottomrule[0.75pt]
\end{tabular}
\label{analysis}
\end{table}

\begin{figure*}[t]
\centering
\subfigure[]{
\includegraphics[width=0.4\linewidth]{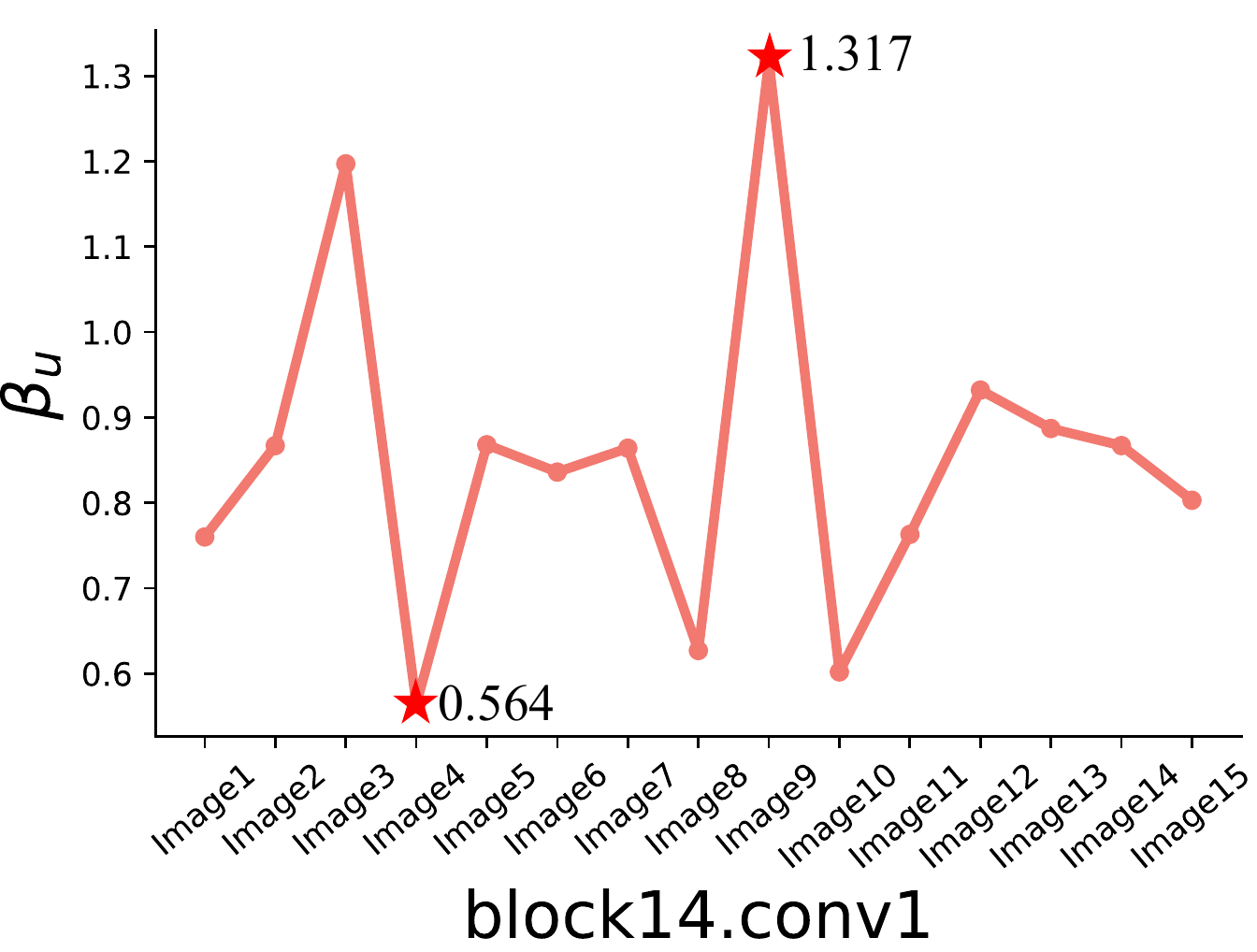} \label{gate:effect}
}
\subfigure[]{
\includegraphics[width=0.4\linewidth]{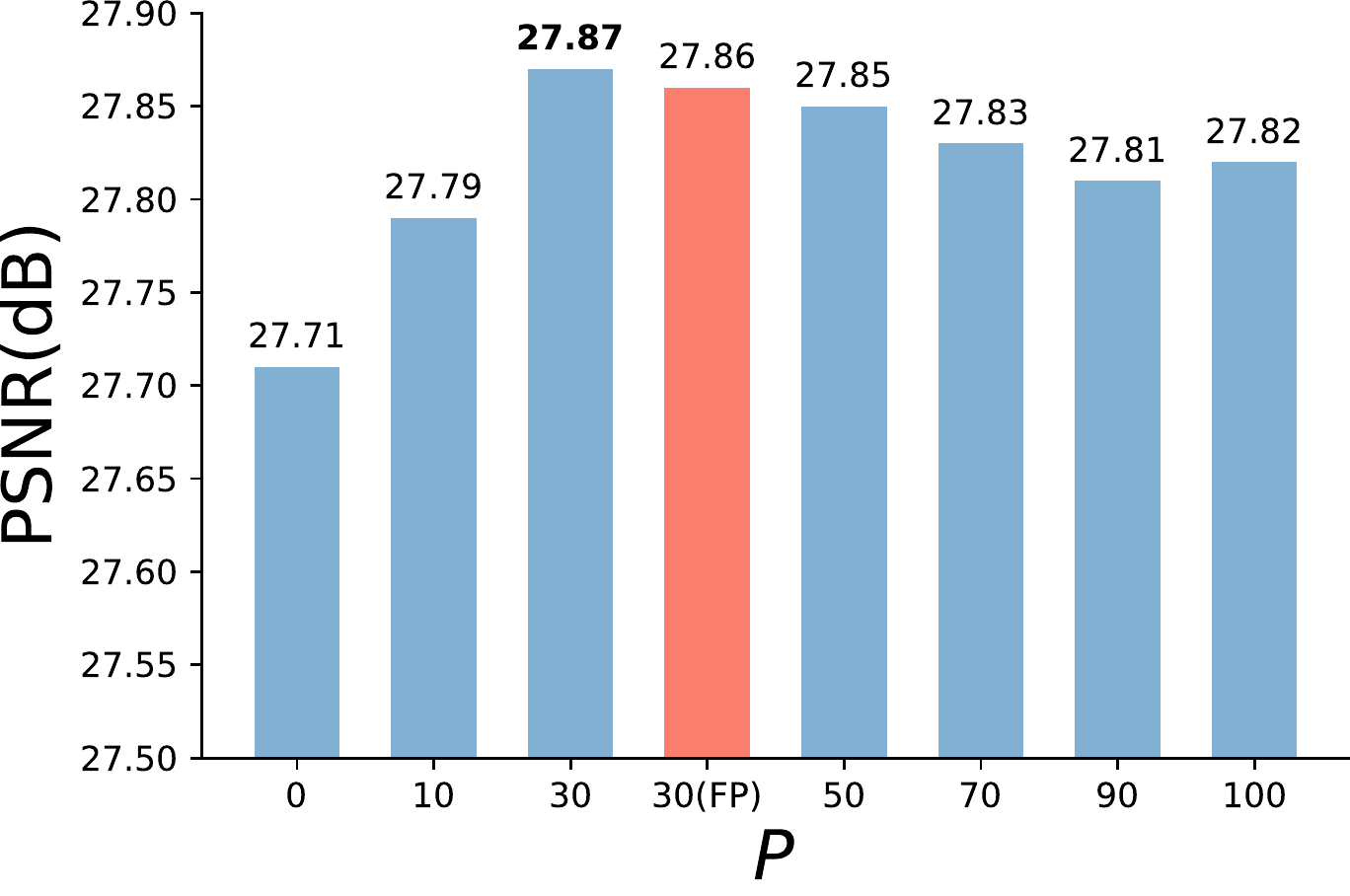}
\label{gate:ratio} 
}
\vspace{-1.0em}
\caption{(a) The $\beta_u$ in 2-bit EDSR$\times$4. (b) Influence of the gate ratio $P$ in 2-bit EDSR$\times$4.}
\label{gate}
\end{figure*}

\textbf{Gate Effect}. Fig.\,\ref{gate:effect} displays the $\beta_u$ values of 15 randomly selected test images. We select the results of 2-bit EDSR$\times$4. It can be seen that the $\beta_u$ of different images varies a lot, which proves that the dynamic gate controller can well adjust the clipping bounds adaptive to the input images.

\subsection{Ablation Study}

This section conducts the ablation study of our DDTB. All experiments are conducted with 2-bit EDSR$\times$4.

\textbf{Gate Parameters}. Fig.\,\ref{gate:ratio} exhibits the results of different gate ratio $P$ on Set14~\cite{ledig2017photo} dataset. The best result is observed when the $P$ is set to 30. Compared with the case without our gate controller, its PSNR increases by 0.16dB. Using more layers cannot bring improvements and even does slight damage to the performance. 
Moreover, using the full-precision gate does not provide better results, which indicates that a 2-bit gate is sufficient.
To reduce search overhead, we find the best $P$ on the model of scale 4 and apply it to the corresponding model of scale 2. Though not optimal, it is sufficient to show SOTA performance.

\begin{table}[!t]
\setlength\tabcolsep{4pt}
\centering
\caption{Effect of different components in our methods. ``DW'': dual bounds for weight quantization. ``DI'': DDTB initializer. The PSNR/SSIM are reported.}
\vspace{-1.0em}
\begin{tabular}{ccc|cccc} \toprule[1.25pt]
\multicolumn{3}{c|}{Components} & \multicolumn{4}{c}{Results}       \\ \hline \hline
DDTB       & DW       & DI      & Set5~\cite{bevilacqua2012low}  & Set14~\cite{ledig2017photo} & BSD100~\cite{martin2001database} & Urban100~\cite{huang2015single} \\ \hline
\multicolumn{3}{c|}{PAMS~\cite{li2020pams}}       & 29.51/0.835 & 26.79/0.734 & 26.45/0.696   &  23.72/0.688    \\ \hline
\checkmark   &  &  & 30.00/0.853 & 27.15/0.745 & 26.69/0.705 & 24.06/0.707  \\
   &  \checkmark    &    & 29.82/0.846& 27.02/0.741 &  26.62/0.702 & 23.98/0.702 \\
     &  \checkmark  &  \checkmark    & 30.19/0.852
 & 27.30/0.747  & 26.76/0.703 & 24.19/0.711\\
\checkmark   &          &  \checkmark & 30.23/0.858 & 27.33/0.750 &  26.78/0.709 & 24.23/0.714 \\
   \checkmark   &  \checkmark  &         & 30.80/0.871 & 27.68/0.760  & 26.99/0.717 & 24.64/0.735\\
\checkmark   &   \checkmark  &   \checkmark  & \textbf{30.97/0.876}  & \textbf{27.87/0.764} & \textbf{27.09/0.719}  &  \textbf{24.82/0.742} \\\bottomrule[0.75pt]       
\end{tabular}
\label{ablation:components}
\end{table}

\textbf{Components}. 
We use PAMS as the baseline to show the effect of different components including dynamic dual training bounds (DDTB) for activation quantization, dual bounds for weight quantization, and DDTB initializer (DI).
Table\,\ref{ablation:components} shows the experimental results.
When DDTB and DW are individually added, the performance increases compared with the baseline. The DDTB significantly boosts the baseline which proves its ability to fit the asymmetric activation distribution. By combining the initializer, the performance continues to increase. When all of them are applied, the best performance can be obtained.

\section{Conclusion}
This paper presents a novel quantization method, termed Dynamic Dual Trainable Bounds (DDTB) to solve the asymmetric activation distribution in the DCNN-based SR model. Our DDTB introduces trainable upper and lower bounds, to which a dynamic gate controller is applied in order to adapt to the input sample at runtime. 
The gate is represented in a 2-bit format and only applied to part of the network to minimize the extra overhead. Moreover, we design a special DDTB initializer for stable training. Our DDTB shows its superiority over many competitors with different quantized SR models on many benchmarks, especially when performing ultra-low precision quantization.

\section*{Acknowledgement}
\begin{sloppypar}
This work is supported by the National Science Fund for Distinguished Young Scholars (No. 62025603), the National Natural Science Foundation of China (No. U1705262, No. 62176222, No. 62176223, No. 62176226, No. 62072386, No. 62072387, No. 62072389, No. 62002305, No. 61772443, No. 61802324 and No. 61702136),
Guangdong Basic and Applied Basic Research Foundation (No.2019B1515120049), the Natural Science Foundation of Fujian Province of China (No. 2021J01002), and the Fundamental Research Funds for the central universities (No. 20720200077, No. 20720200090 and No. 20720200091).
\end{sloppypar}

%
%
\bibliographystyle{splncs04}
\bibliography{egbib}

\clearpage

\section*{Appendix \label{appendix}}

\section{Model Complexity}

Table\,\ref{supp:analysis} provides the complexity analyses of 4, 3, 2-bit SR models. The extra overhead of the dynamic gate controller is negligible.

\begin{table}[ht]
\setlength\tabcolsep{5pt}
\scriptsize
\centering
\caption{Complexity analysis. The number in brackets indicates the parameters in the high-level feature extraction module. We compute BOPs by generating a 1920$\times$1080 image (upscaling factor $\times$4).}
\begin{tabular}{ccccccc}
\toprule[1.25pt]
Model & Bit & Params & Gate Params($ratio$) & BOPs & Gate BOPs($ratio$)  \\  \hline \hline
EDSR  & 32  &    1.52M   &   0      &  532T   & 0          \\
EDSR\_DDTB & 2  &  0.41M(0.08M)    &   0.6\%  & 219T       &  0.0000013\%  \\
EDSR\_DDTB & 3  &  0.45M(0.11M)    &  0.6\%   & 220T   &   0.0000013\%  \\
EDSR\_DDTB & 4  &   0.49M(0.15M)    & 0.5\%        &  222T  & 0.0000013\% \\ \hline
RDN  & 32  &    22.3M  &   0      & 6038T  &    0      \\
RDN\_DDTB & 2  &  1.76M(1.42M)       &    2.8\%         &  239T &  0.0000066\%  \\
RDN\_DDTB & 3  & 2.44M(2.10M)     &   2\%   & 267T & 0.0000059\%  \\
RDN\_DDTB & 4  &  3.13M(2.79M)  &    1.6\%     &   307T & 0.0000051\%\\ \hline
SRResNet  & 32  &    1.543M  &   0      & 591T  &    0       \\
SRResNet\_DDTB & 2  & 0.44M(0.07M)     &   0.1\%    & 278T  &  0.0000002\%    \\
SRResNet\_DDTB & 3  &  0.47M(0.11M)   &   0.1\%   &  280T  &  0.0000002\%      \\
SRResNet\_DDTB & 4  &  0.51M(0.15M)&  0.1\%   &  282T  & 0.0000002\% \\
\bottomrule[0.75pt]
\end{tabular}
\label{supp:analysis}
\end{table}

\clearpage
\section{Results of Fully Quantized Models}

In this section, we provide the comparisons between the existing fully quantized method FQSR~\cite{wang2021fully} and our DDTB. Following FQSR~\cite{wang2021fully}, all layers and skip-connections of SR models are quantized. As shown in Table\,\ref{full-quantized}, DDTB outperforms FQSR by a large margin when performing 4-bit quantization. For instance, DDTB obtains performance gains by 0.98dB, 0.58dB, 0.37dB, and 0.77dB on Set5, Set14, BSD100, and Urban100, respectively.

\begin{table}[ht]
\scriptsize
\setlength\tabcolsep{3pt}
\centering
\caption{PSNR/SSIM comparisons between the existing fully quantized method and our DDTB. ``SC'' indicates the bit-width of skip-connections.}
\begin{tabular}{cccccccc}
\toprule[1.25pt]
Model                                                               & Bit                   & SC & Methods        & Set5       & Set14          & BSD100          & Urban100          \\ \hline \hline
\multirow{2}{*}{\begin{tabular}[c]{@{}c@{}}EDSR\\ $\times$4\end{tabular}} & \multirow{2}{*}{\begin{tabular}[c]{@{}c@{}}4\end{tabular}}  & \multirow{2}{*}{\begin{tabular}[c]{@{}c@{}}8\end{tabular}}  & FQSR & 30.93/0.870 &  27.82/0.761 & 27.07/0.715 &  24.93/0.744 \\
                                                                    &                     &          & \textbf{DDTB(Ours)}        & \textbf{31.91/0.889} & \textbf{28.40/0.777} & \textbf{27.44/0.732} & \textbf{25.70/0.775}        
                                                                    \\\hline
\multirow{2}{*}{\begin{tabular}[c]{@{}c@{}}EDSR\\ $\times$2\end{tabular}} & \multirow{2}{*}{\begin{tabular}[c]{@{}c@{}}4\end{tabular}}  & \multirow{2}{*}{\begin{tabular}[c]{@{}c@{}}8\end{tabular}}  & FQSR & 37.04/0.951 &   32.84/0.908 &  31.67/0.889 &   30.65/0.911 \\
                                                                    &                     &          & \textbf{DDTB(Ours)}        & \textbf{37.83/0.960} & \textbf{33.44/0.916} & \textbf{32.07/0.898} & \textbf{31.60/0.924}        
                                                                    \\\hline
\multirow{2}{*}{\begin{tabular}[c]{@{}c@{}}SRGAN\\ $\times$4\end{tabular}} & \multirow{2}{*}{\begin{tabular}[c]{@{}c@{}}4\end{tabular}}  & \multirow{2}{*}{\begin{tabular}[c]{@{}c@{}}8\end{tabular}}  & FQSR & 30.96/0.872 &  27.85/0.759 & 27.08/0.713 & 24.93/0.742 \\
                                                                    &                     &          & \textbf{DDTB(Ours)}        & \textbf{31.46/0.879} & \textbf{28.10/0.766} & \textbf{27.26/0.721} & \textbf{25.33/0.757}         
                                                                    \\\hline
\multirow{2}{*}{\begin{tabular}[c]{@{}c@{}}SRGAN\\ $\times$2\end{tabular}} & \multirow{2}{*}{\begin{tabular}[c]{@{}c@{}}4\end{tabular}}  & \multirow{2}{*}{\begin{tabular}[c]{@{}c@{}}8\end{tabular}}  & FQSR & 36.69/0.950 &  32.64/0.906  & 31.57/0.888 & 30.37/0.908 \\
                                                                    &                     &          & \textbf{DDTB(Ours)}        & \textbf{36.84/0.950} & \textbf{32.79/0.907} & \textbf{31.69/0.889} & \textbf{30.70/0.910}         
                                                                    \\\hline
\multirow{2}{*}{\begin{tabular}[c]{@{}c@{}}SRResNet\\ $\times$4\end{tabular}} & \multirow{2}{*}{\begin{tabular}[c]{@{}c@{}}4\end{tabular}}  & \multirow{2}{*}{\begin{tabular}[c]{@{}c@{}}8\end{tabular}}  & FQSR & 31.04/0.874 & 27.86/0.761 & 27.09/0.714 & 24.95/0.744 \\
                                                                    &                     &          & \textbf{DDTB(Ours)}        & \textbf{31.51/0.881} & \textbf{28.17/0.767} & \textbf{27.31/0.722} & \textbf{25.39/0.760}         
                                                                    \\\hline
\multirow{2}{*}{\begin{tabular}[c]{@{}c@{}}SRResNet\\ $\times$2\end{tabular}} & \multirow{2}{*}{\begin{tabular}[c]{@{}c@{}}4\end{tabular}}  & \multirow{2}{*}{\begin{tabular}[c]{@{}c@{}}8\end{tabular}}  & FQSR & 36.34/0.945 & 32.40/0.901 &  31.37/0.882 & 29.98/0.899 \\
                                                                    &                     &          & \textbf{DDTB(Ours)}        & \textbf{36.85/0.951} & \textbf{32.73/0.907} & \textbf{31.63/0.890} & \textbf{30.63/0.910}         
                                                                    \\\bottomrule[0.75pt]
\end{tabular}
\label{full-quantized}
\end{table}

\clearpage
\section{Visualization}

\begin{figure*}[ht]
\centering
\subfigure[2-bit EDSR$\times$2.]{
\includegraphics[width=0.9\linewidth]{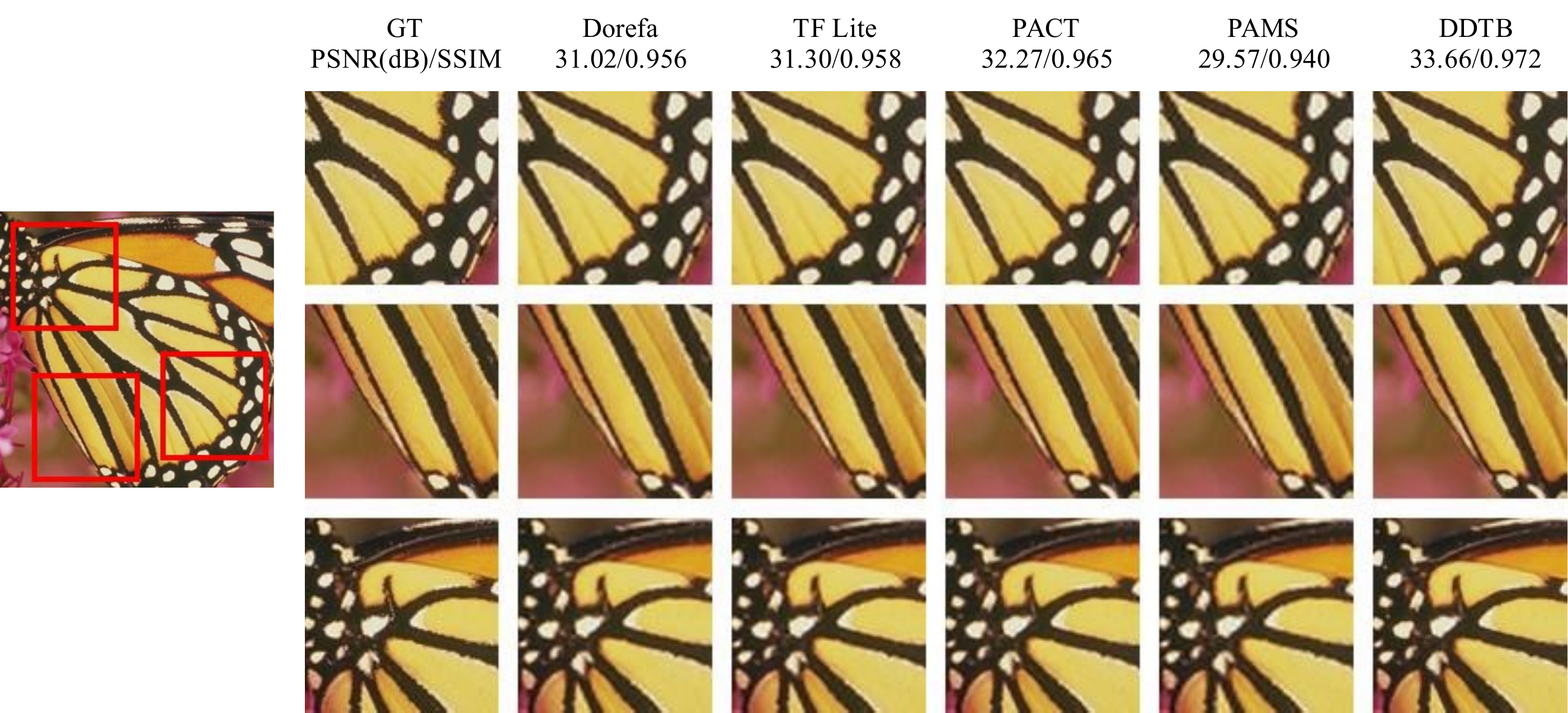} \label{supp:vis-2bit-edsrx2}
}
\\
\subfigure[2-bit EDSR$\times$4.]{
\includegraphics[width=0.9\linewidth]{vis-2bit-edsrx4.pdf} \label{supp:vis-2bit-edsrx4}
}
\caption{Reconstructed results of 2-bit EDSR$\times$2 and EDSR$\times$4.}
\label{supp:vis-edsr}
\end{figure*}

\begin{figure*}[t]
\centering
\subfigure[2-bit RDN$\times$2.]{
\includegraphics[width=0.9\linewidth]{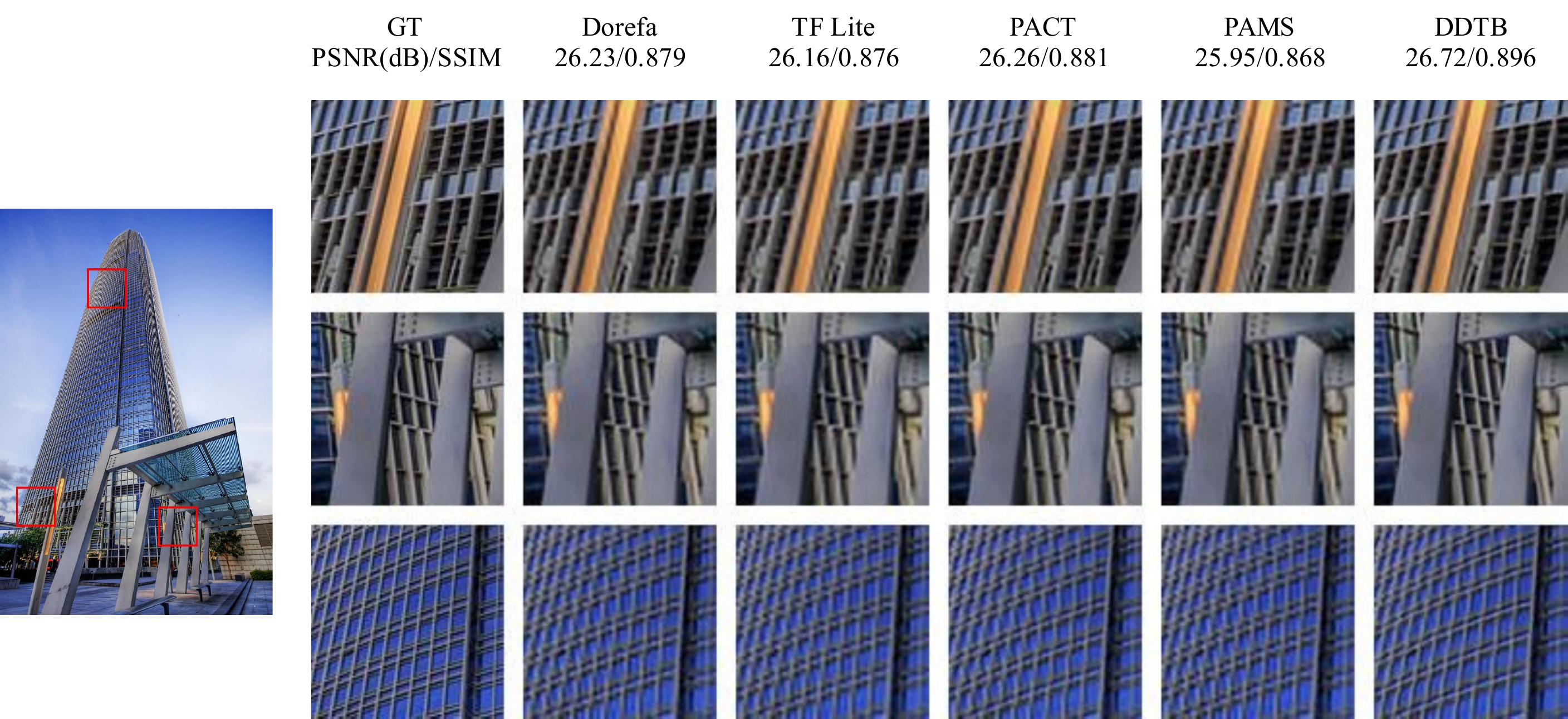} \label{supp:vis-2bit-rdnx2}
}
\\
\subfigure[2-bit RDN$\times$4.]{
\includegraphics[width=0.9\linewidth]{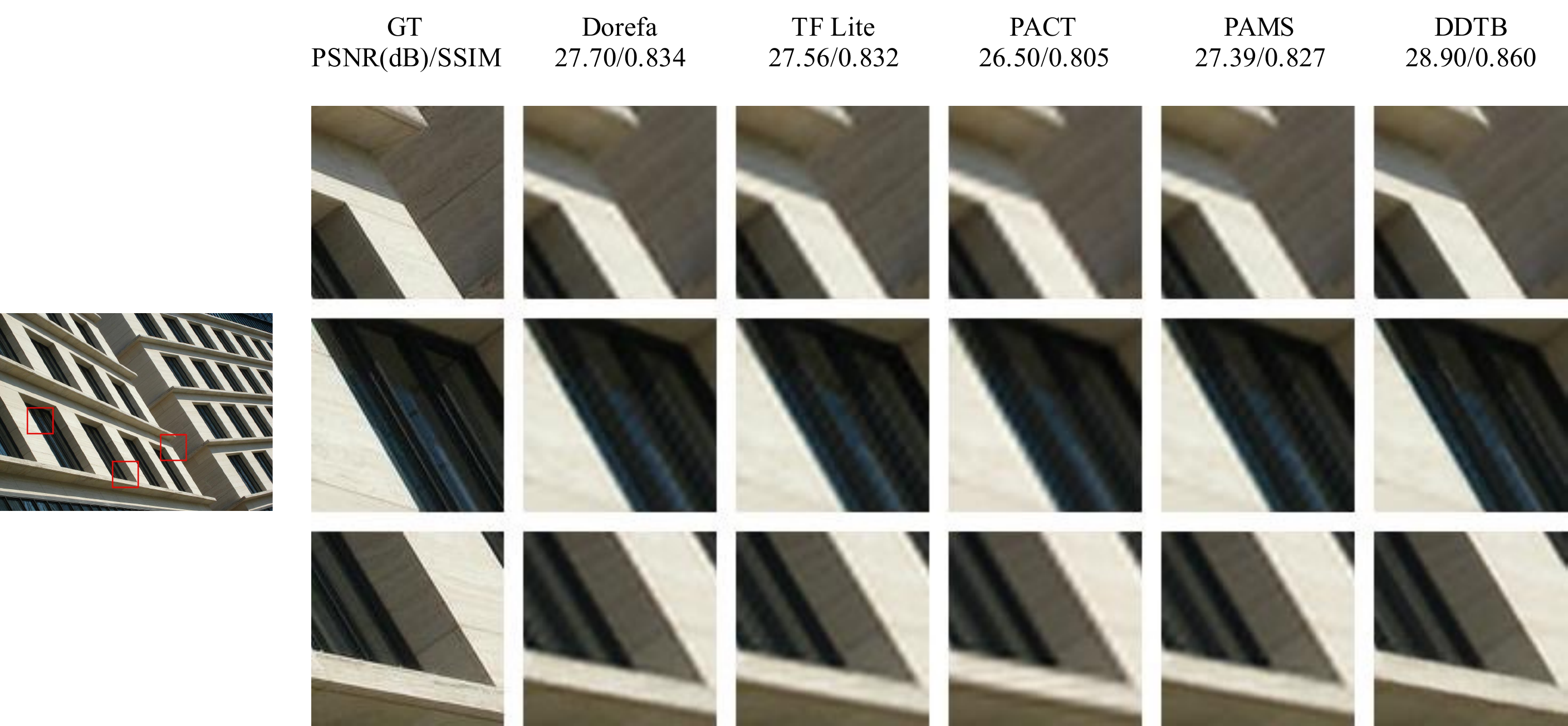} \label{supp:vis-2bit-rdnx4}
}
\caption{Reconstructed results of 2-bit RDN$\times$2 and RDN$\times$4.}
\label{supp:vis-rdn}
\end{figure*}

\begin{figure*}[t]
\centering
\subfigure[2-bit SRResNet$\times$2.]{
\includegraphics[width=0.9\linewidth]{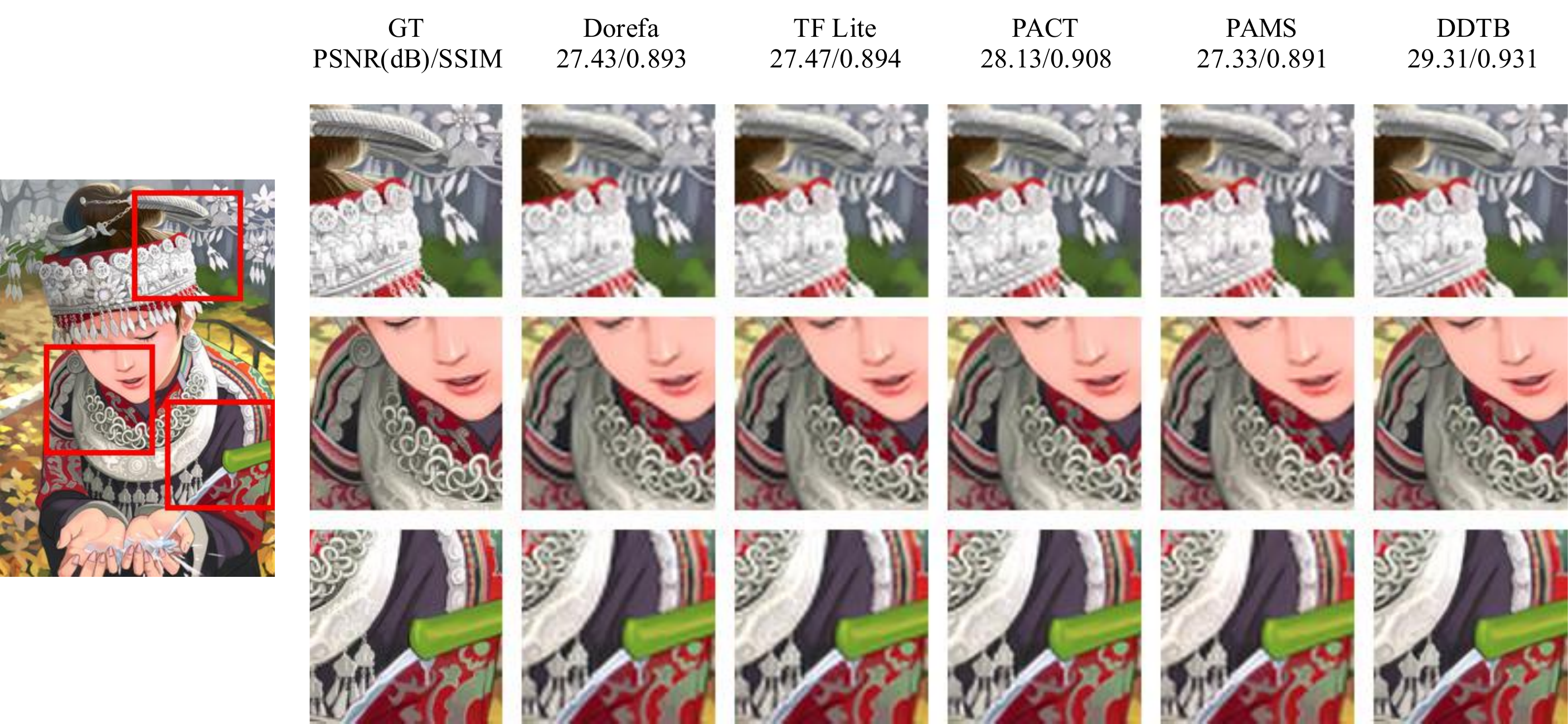} \label{supp:vis-2bit-srrx2}
}
\\
\subfigure[2-bit SRResNet$\times$4.]{
\includegraphics[width=0.9\linewidth]{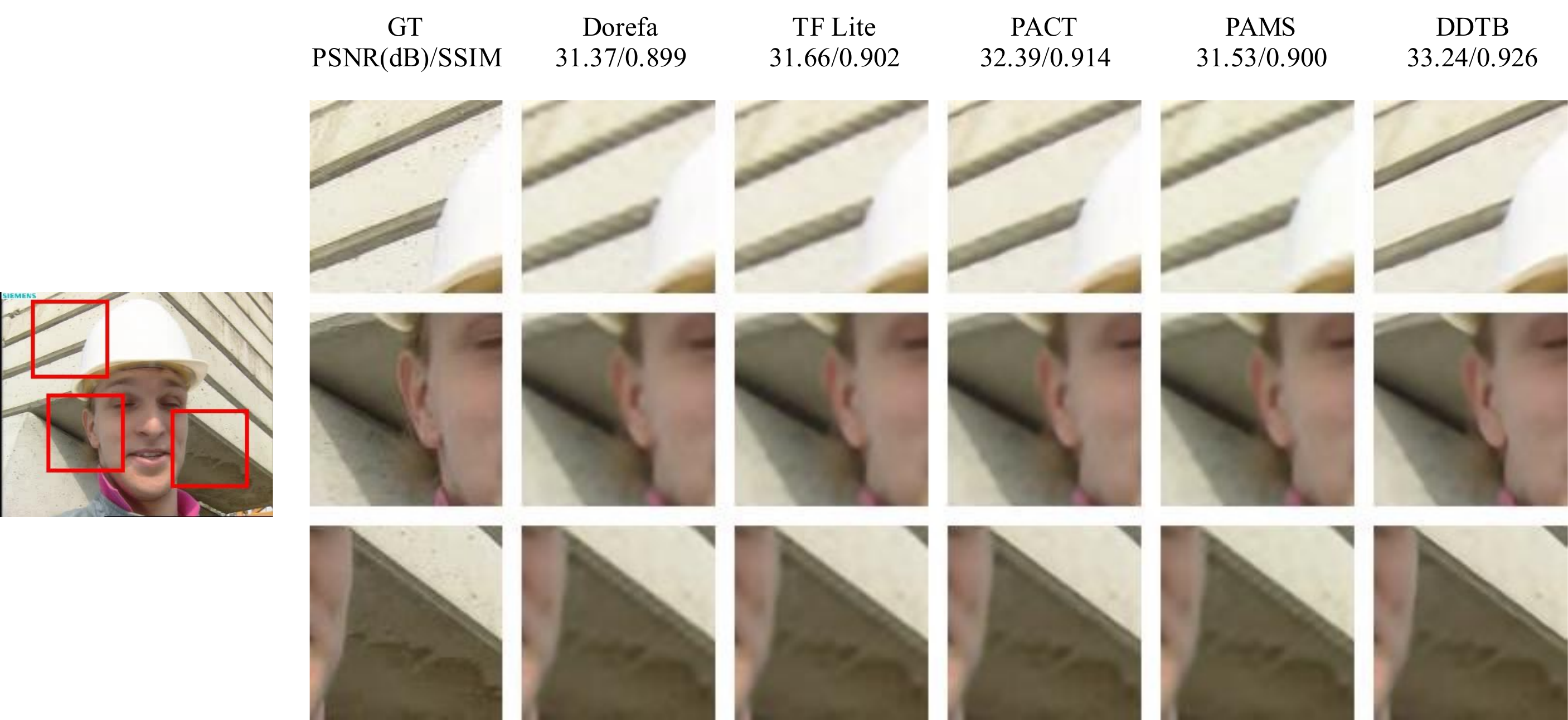} \label{supp:vis-2bit-srrx4}
}
\caption{Reconstructed results of 2-bit SRResNet$\times$2 and SRResNet$\times$4.}
\label{supp:vis-srr}
\end{figure*}

Fig.\,\ref{supp:vis-edsr}, Fig.\,\ref{supp:vis-rdn}, and Fig.\,\ref{supp:vis-srr} exhibit the reconstructed results of the 2-bit EDSR, 2-bit RDN, and 2-bit SRResNet, respectively. The reported PSNR/SSIM are measured by the displayed image.

\end{document}